# Imaging of super-fast dynamics and flow instabilities of superconducting vortices

L. Embon[1*], Y. Anahory[1,2*], Ž.L. Jelić[3,4*], E.O. Lachman[1], Y. Myasoedov[1], M.E. Huber[5], G.P. Mikitik[6], A.V. Silhanek[4], M.V. Milošević[3], A. Gurevich[7], E. Zeldov[1]

[1]Department of Condensed Matter Physics, Weizmann Institute of Science, Rehovot 7610001, Israel

[2]Racah Institute of Physics, The Hebrew University, Jerusalem 91904, Israel

[3]Departement Fysica, Universiteit Antwerpen, Groenenborgerlaan 171, B-2020, Antwerpen, Belgium

[4]Département de Physique, Université de Liège, B-4000 Sart Tilman, Belgium

[5]Departments of Physics and Electrical Engineering, University of Colorado Denver, Denver 80217, USA

[6]Verkin Institute for Low Temperature Physics & Engineering, Ukrainian Academy of Sciences, Kharkov 61103, Ukraine

[7]Department of Physics, Old Dominion University, Norfolk, VA 23529-0116, USA

*these authors had equal contribution

Correspondence to: yonathan.anahory@mail.huji.ac.il eli.zeldov@weizmann.ac.il

**Quantized magnetic vortices driven by electric current determine key electromagnetic properties of superconductors. While the dynamic behavior of slow vortices has been thoroughly investigated, the physics of ultrafast vortices under strong currents remains largely unexplored. Here we use a nanoscale scanning superconducting quantum interference device to image vortices penetrating into a superconducting Pb film at rates of tens of GHz and moving with velocities up to tens of km/s, which are not only much larger than the speed of sound but also exceed the pair-breaking speed limit of superconducting condensate. These experiments reveal formation of mesoscopic vortex channels which undergo cascades of bifurcations as the current and magnetic field increase. Our numerical simulations predict metamorphosis of fast Abrikosov vortices into mixed Abrikosov-Josephson vortices at even higher velocities. This work offers an insight into the fundamental physics of dynamic vortex states of superconductors at high current densities, crucial for many applications.**

The dynamics of current-driven vortex matter is of major importance both for the comprehension of the fundamental collective behavior of strongly interacting vortices and for attaining high non-dissipative currents in superconductors for applications. Materials advances in incorporating artificial pinning centers that immobilize vortices, particularly oxide nano-precipitates in cuprates, have resulted in critical current densities $J_c$ as high as 10-20% of the depairing current density $J_d$ at which the superconducting state breaks down[1–3]. At such high current densities $J$, once a vortex gets depinned from a defect, it can move with high velocity $v$ and dissipate much power. Understanding this phenomenon is critical for many applications, such as high-field magnets[4], superconducting digital memory and qubits[5], THz radiation sources[6], or resonator cavities for particle accelerators[7]. Yet, little is known about what happens to a vortex driven by very strong currents at the depairing limit $J \sim J_d$ and what is the maximal terminal velocity a vortex can reach. Moreover, even a more fundamental question of whether the notion of a moving vortex as a stable topological defect[8,9] remains applicable at ultrahigh velocities has not been explored.

At high current densities with $J \gg J_c$ the effect of the disorder-induced pinning force diminishes and the resulting velocity of a vortex $v$ is mainly determined by the balance of the driving Lorentz force per vortex unit length $F_L = \phi_0 J$ and the viscous drag force, $F_d = \eta(v) v$ [8–10]. At small $v$ the viscous Bardeen-Stephen drag coefficient, $\eta_0 \simeq \phi_0^2/2\pi\xi^2 \rho_n$, results from dissipation in a circular, non-superconducting vortex core of radius $\simeq \xi$. Here the coherence length $\xi = \hbar v_F/\pi\Delta$ defines the size of the Cooper pair in the clean limit, $\Delta$ is the superconducting gap at $J = 0$, $v_F$ is the Fermi velocity, $\rho_n$ is the normal state electrical resistivity and $\phi_0 = 2.07 \cdot 10^{-15}$ Wb is the magnetic flux quantum[8,9]. Since the current density is limited by the depairing value $J_d \simeq \phi_0/4\pi\mu_0\lambda^2\xi$ at which the speed of the superconducting condensate reaches the pair-breaking velocity $v_{dp} = \Delta/mv_F = \hbar/\pi m\xi$ [9], the maximal vortex velocity can be extrapolated to $v_c \simeq \phi_0 J_d/\eta_0 \simeq \rho_n \xi/2\mu_0\lambda^2$ where $\lambda$ is the magnetic penetration depth and $m$ is the effective electron mass[7]. For a Pb film with $\lambda = 96$ nm and $\xi = 46$ nm at $T = 4.2$ K, and $\rho_n \sim 20$ nΩm [11], we obtain $v_{dp} \simeq 0.4$ km/s and $v_c \sim 40$ km/s, which suggests that the vortex could move at a velocity that is two orders of magnitude higher than the maximal drift velocity of the Cooper-pair condensate. A vortex moving much faster than the perpendicular current superflow which drives it raises many fundamental issues. What is the maximal terminal velocity that a single vortex can actually reach and what are the mechanisms that set this limit? Does a vortex remain a well-defined topological defect even under the extreme conditions of the strongest possible current drive? Does the superfast vortex matter form dynamic patterns qualitatively different from the conventional flux flow at low velocities? Some of these issues have been studied in numerical simulations of the time-dependent Ginzburg-Landau (TDGL) equations, which, however, are only applicable at temperatures very close to $T_c$ [12–16]. Since suitable theoretical frameworks for exploring the extreme dynamics of superfast vortices at low temperature have not yet been developed (see Supplementary Note 1), the role of the experiment becomes paramount.

Addressing the physics of fast vortices experimentally is extremely challenging. For instance, inferring the terminal velocity $v_c$ from the conventional measurements of dc voltage-current (V-I) characteristics[16–19] is rather indirect because it assumes that all vortices move with the same constant velocity, which is not the case, as will be shown below. Therefore, a local probe capable of tracing vortices moving at supersonic velocities is required. A number of methods, including STM[20–22], MFM[23,24], magneto-optical imaging[25], scanning superconducting quantum interference device (SQUID) microscopy[26], and scanning Hall probes[27–29] have been employed to image slowly moving vortex structures, but none of them could resolve the properties of high-speed vortices. In this work we employ a novel SQUID that resides on the apex of a sharp tip[30] and provides high spatial-resolution magnetic imaging[31,32] reaching single-spin sensitivity[30] and enabling detection of sub-nanometer ac vortex displacements[11]. Using this nanoscale SQUID-on-tip (SOT) we report the first direct microscopic imaging of superfast vortices under current densities approaching the depairing limit. Our experiment revealed vortex velocities up to tens of km/s, cascades of striking branching instabilities, and dynamic transitions in the moving vortex matter. Comprehension of the fundamental vortex properties under these extreme, previously unexplored conditions is essential for reducing dissipation and preventing breakdown of superconductivity in high current applications.

## Results

**Imaging of stationary and flowing vortices.** Our experiments were performed on a Pb film with thickness $d = 75$ nm and $T_c = 7.2$ K patterned into a 10 μm-wide microbridge with a central constriction of width $w = 5.7$ μm (Fig. 1, Supplementary Note 2, and Supplementary Figs. 1 and 2). In this geometry vortices only penetrate in the narrowest part of the bridge, which greatly reduces heating. Imaging of the local magnetic field $B_z(x,y)$ above the film surface at 4.2 K was done using a 228 nm diameter SOT incorporated into a scanning probe microscope (see Methods). Figures 2a-d show the distribution of vortices in the strip after field cooling in $B_a = 2.7$, 5.4, and 9.0 mT which display a disordered vortex structure pinned by material defects. The observed vortex density is not uniform, as one may expect under field cooling conditions, but has a dome-shaped profile with a maximum in the center surrounded by vortex-free stripes along the edges. This is the result of the geometrical barrier[33] which is strikingly demonstrated here with single-vortex resolution (see also Supplementary Fig. 3). Unlike a bulk superconductor in which the screening currents flow in a narrow layer of thickness $\lambda$ at the surface, in a thin film strip of width $0 < x < w$ and thickness $d \ll w$ in perpendicular field $B_a$, the shielding current density in the Meissner state $J_M(x) = B_a(w - 2x)/(d\mu_0\sqrt{x(w-x)})$ varies over much larger scales and decreases slowly $J_M(x) \propto x^{-1/2}$ and $J_M(x) \propto -(w-x)^{-1/2}$ away from the left and right edges, respectively (see Fig. 1b). These currents push vortices into the central part of the strip, where they form a magnetic flux dome surrounded by vortex-free regions[33–35]. The vortex-free region shrinks with $B_a$ as seen in Fig. 2.

These vortex-free regions have a major effect on vortex dynamics in the presence of transport current. As the applied current $I$ is increased, the sum of transport and shielding current densities $J(x)$ shown in Fig 1b increases at the left edge ($x = 0$) and decreases at the right edge ($x = w$). As a result, the vortex dome shifts towards the right edge and the vortex-free region at the left edge expands[33,36], as shown in Figs. 2e-h at $I \lesssim I_c$. At the critical current $I = I_c$, the current density at the left edge approaches the depairing limit $J(0) \cong J_d$, and the flux dome reaches the right edge where $J(x)$ vanishes, so that the conditions for the onset of vortex motion are met. Here the critical state revealed with a single vortex resolution, is dominated by the geometrical and extended surface barriers[33–37], which has two essential differences as compared to the continuum, pinning-dominated Bean critical state[38,39]. First, unlike the Bean state in which the vortex density is highest at the penetration edge of the sample, Figs. 2e-h show zero vortex density at the penetration side (left). Secondly and most importantly, in the Bean model at $I = I_c$ the current density equals $J_c$ across the entire sample, whereas our thin film bridge is separated into two distinct regions clearly seen in Fig. 2f. In the left vortex free region, $J$ significantly exceeds the critical current density, $J_c < J < J_d$, and no stationary vortices can be present[33–36]. In the right half where $0 < J \leq J_c$ vortices are pinned. It is this unique inhomogeneous current state which allows us to investigate dynamics of superfast vortices driven by high local current densities that cannot be done by global transport measurement in bulk samples. Here the penetrating vortices can be subjected locally to extremely high current densities $J \gg J_c$ at the edges while the net current is only slightly above the critical, $I \gtrsim I_c$ and heating is weak.

As $I$ exceeds $I_c$, vortices start penetrating through the left edge of the constriction, as shown in Figs. 2i-l. Since vortices traverse the sample in about 1 ns, much shorter than our imaging time (~4 min/image), the observed images represent time-averaged locations of vortices. Remarkably, rather than entering randomly at the edge and flowing across the film while avoiding each other due to repulsive interactions, vortices penetrate at a well-defined point in the narrowest part of the bridge and follow each other forming a single channel or stem, which then undergoes subsequent bifurcations. The number of stems and the number of branches increases with increasing $I$ and $B_a$, as shown in Figs. 2i-p (see Supplementary Movies 1-4). The intensity of the local $B_z(x,y)$ along the channels is proportional to the fraction of time spent by a vortex at a specific location. The variations in $B_z(x,y)$ with distinct maxima along some of the channels (see, e.g., Fig. 2i) thus reveal the locations where the vortices slow down due to pinning and vortex-vortex interlocking. The field profiles along the channels become more uniform as the current increases (see Fig. 2m).

**Transport properties and vortex penetration frequency.** In order to extract the vortex velocity, we measured the $V - I$ characteristics simultaneously with the SOT imaging, as shown in Fig. 3a. At each field, the onset of finite voltage $V$ coincides with the appearance of the first vortex channel as $I$ exceeds a critical current $I_c(B_a)$ that decreases with $B_a$. The data points end at the maximal current above which the voltage jumps abruptly by more than an order of magnitude (black arrows) apparently due to a thermal quench in the constriction region. Such simultaneous SOT and transport measurements provide a unique opportunity to reveal superfast dynamics of vortices and their trajectories with a single-vortex resolution. Figure 3b shows the measured voltage drop on the bridge (left axis) along with the number of vortex stems $n$ observed by SOT imaging (right axis) vs. current at $B_a = 2.7$ mT (see Movie 1). The appearance of each subsequent stem in Fig. 3b matches a step in the voltage and a change in the differential resistance $dV/dI$. Linear fits to the data (dashed) show a roughly two-fold increase in $dV/dI$, from 13.9 mΩ for one stem to 25.1 mΩ for two stems. For a given number of stems $n$, the vortex penetration frequency $f$ in each stem is given by the Faraday law, $f = V/n\phi_0$. Figure 3c shows that the penetration frequency jumps from zero to 3.7 GHz at the formation of the first stem. As $I$ increases, $f$ rises to 15.3 GHz and then drops abruptly to 9.1 GHz upon the formation of the second stem.

**Vortex velocity.** Vortex conservation requires that $f = V/\phi_0$ remains constant along the stem up to the bifurcation point, so that the vortex velocity along the stem is given by $v(x) = fa(x)$ where $a(x)$ is the local average intervortex distance. Our simultaneous SOT and transport measurements allow determination of $a(x)$ and $v(x)$ as follows. The average field along a chain of vortices separated by $a(x)$ is given by $B_{av}(x) = \int_{-\infty}^{\infty} B_v(u)du/a(x)$, where $B_v(u)$ is the magnetic field profile of an individual vortex. By measuring $B_{av}(x)$ along the stem and $B_v(x)$ across an isolated stationary vortex (see Supplementary Note 3 and Supplementary Fig. 4), we thus obtain $a(x)$ along a single stem (Fig. 4a) from $x = 0.5$ μm up to the bifurcation point $x = x_b$, for various currents at $B_a = 2.7$ mT. Taking the penetration rate $f$ from Fig. 3c, we derive the corresponding vortex velocity $v(x) = fa(x)$ as shown in Fig. 4b. The resulting $v(x)$ decreases with the distance $x$ from the edge, consistent with the decreasing current density near the edge[33–36], $J(x) \simeq J_d(\Lambda/x)^{1/2}$, which drives the vortices, where $\Lambda = 2\lambda^2/d$. The remarkable findings here are the extreme values of vortex velocities of 10-20 km/s, much larger than the depairing superfluid velocity $v_d \simeq 0.4$ km/s estimated above.

These very high velocities are attained at 0.5 μm $< x <$ 1.5 μm where the estimated current density is $0.6J_d > J > 0.2J_d$. In the region of $0 < x < 0.5$ μm at the film edge, where even higher currents and vortex velocities are possible, our SOT technique cannot resolve the actual $v(x)$ as the vortex field $B_v(x)$ close to the edge becomes partly extinguished by the image vortex imposed by the boundary conditions[40] (see Supplementary Note 3). Figure 4c shows that the vortex velocities at $x = 0.5$ μm and at $x = x_b$ increase monotonically with $I$, whereas the length of the stem defined by the bifurcation point $x_b$ decreases with $I$, as seen in Fig. 4b.

The average vortex velocity in the stem can be estimated independently of the above analysis by assuming the distance between the moving vortices to be of the order of their mean stationary distance $a \cong 1$ μm from Fig. 2b (which is close to $a = \left(2\phi_0/\sqrt{3}B_a\right)^{1/2} = 0.94$ μm) and taking the highest frequency $f \cong 15$ GHz from Fig. 3c. This yields $v = fa \cong 15$ km/s which is consistent with the measured vortex velocities in Fig. 4b.

The mesoscopic chains of single vortices moving along stationary channels under a dc drive and weak overheating reported here are fundamentally different from transient dendritic flux avalanches observed by magneto-optical imaging in increasing magnetic fields[41–45]. Those macroscopic filaments of magnetic flux focused in regions overheated above $T_c$ can propagate with velocities as high as 150 km/s in $YBa_2Cu_3O_7$ films at 10 K[43] or 360 km/s in $YNi_2B_2C$ at 4.6 K[44]. Such thermomagnetic avalanches are driven not by the motion of single vortices but by strong inductive overheating caused by the fast-propagating stray electromagnetic fields outside the film[45], unlike the correlated flow of quantized vortices reveled by our SOT imaging under nearly isothermal conditions. The mechanisms of channeling and branching of fast vortices in our viscosity-dominated regime at $J \gg J_c$ are also different from the disorder-driven formation of networks of slower vortices near the depinning transition observed in numerical simulations[46,47].

**Discussion**

Our experimental findings raise many fundamental questions: What are the mechanisms of vortex confinement in the channels? Why does the branching instability occur and what controls the number of vortex stems? What are the mechanisms which determine the terminal velocities of vortices? How does the structure of a vortex evolve at the superfast velocities observed in our experiments? Given the lack of theory to describe vortices under such extreme conditions, we only limit ourselves to a qualitative discussion of essential effects which may help understand our SOT observations.

The stationary pattern of vortex channels shown in Fig. 2 seems counterintuitive, since vortices repel each other and should therefore disperse over the film. Moreover, each stem grows into a tree through a series of subsequent bifurcations but the branches of different trees do not merge. In order to keep vortices within each channel a mechanism for dynamic alignment of fast moving vortices must be present. One such mechanism is that a rapidly moving vortex leaves behind a wake of reduced order parameter which attracts the following vortex. As a result, a confined chain of vortices in a self-induced channel of reduced superfluid density can be formed (Supplementary Note 1), as it was observed previously in numerical TDGL simulations[15,16] at high currents, $J \sim J_d$, and $T \approx T_c$. As discussed below, this mechanism apparently becomes dominant at velocities substantially higher than those accessible in our experiment.

Vortex alignment may also result from a weak quasiparticle overheating: the power $P = \eta v^2$ generated by each vortex produces a channel of enhanced temperature along the moving vortex chain. The resulting bell-shaped temperature distribution $T(y)$ across the channel causes a restoring force $f_r = -s^* dT/dy$ which stabilizes the vortex chain against buckling distortions (see Supplementary Note 4), where $s^*(T)$ is the transport entropy that defines thermoelectric effects in superconductors[7]. The thermal confinement resulting in long-range alignment of vortices is particularly effective at large vortex spacing $a \gg \xi$ relevant to $a \simeq 1 - 2\ \mu m$. The observed spacing along the stem varies from 0.6 to 2 μm (Fig 4a) much larger than $\xi = 46$ nm, which suggests that the thermal confinement is dominant. As shown in Supplementary Note 4, this mechanism can be effective, even at weak overheating, without causing any thermo-magnetic instabilities.

Once a confined channel is formed, why would it bifurcate? A chain of aligned vortices can become unstable and buckle as repelling vortices get closer to each other. In a thin film, the long-range surface currents produced by vortices result in the repulsion force $f_m = \phi_0^2/\mu_0 \pi a^2$ between vortices spaced by $a \gg 2\lambda^2/d \simeq 250$ nm [8]. As the vortex spacing drops below a critical value $a_c$, the net repulsion force pushing a vortex displaced by $u$ perpendicular to the channel, $f_\perp \sim u\phi_0^2/\mu_0 a^3$, exceeds the restoring force $f_r = -ku$, leading to chain bifurcation, where the spring constant $k(a)$ is determined by the confinement mechanism. The thermal confinement leads to the critical spacing that decreases with the voltage $V$ across the channel as $a_c \propto V^{-1/2}$ (Supplementary Note 4), in qualitative agreement with the experimental data shown in Fig. 4d. As described in Supplementary Note 5, transverse displacements of fast vortices at $J \gg J_c$ can be enhanced by the weak effect of disorder which may cause premature bifurcation of vortex channels.

To gain an insight into the structure and the viscous dynamics and channeling of fast vortices, we performed numerical simulations of TDGL equations for the bridge geometry of Fig. 1 and the material parameters of our Pb films (see Methods). We limit ourselves to a minimal model of superfast vortices driven by strong current densities $J \gg J_c$ for which disorder and heating was neglected. The simulated Cooper pair density $\Delta^2(x, y)$ shown in Fig. 5a reproduces the main features of the SOT images at $I \lesssim I_c$, namely vortices displaced to the right edge and a pronounced vortex-free region along the left edge. Notice that in the absence of disorder, the stationary vortices in Fig. 5a form an ordered structure within a smooth confining potential of the geometrical barrier, in contrast to Fig. 2f which shows a disordered vortex configuration determined by pinning in the right-hand side of the sample where $J < J_c$. At $I = I_c$ the calculated current density $J(x)$ shown in Fig. 5b reaches the depairing limit $J_d$ at the left edge of the constriction and vanishes at the opposite edge.

At $I > I_c$ vortices start penetrating through the left edge and move along a network of preferable paths forming a branching tree with an overall shape determined by the bridge geometry. The vortex chains are curved on larger scales (Fig. 5e) due to the lensing effect of the current distribution in the constriction, which tends to orient the vortex chains perpendicular to the local current $J(x, y)$. The calculated vortex flow pattern is similar to the SOT image in Fig. 2j and also exhibits the coexistence of moving and stationary vortices as observed in Fig. 2 where bulk pinning further hampers the motion of remote vortices. The Copper pair density averaged over the simulated period of time shows a non-uniform distribution along the vortex channels (Figs. 5c,e) with distinct bright spots of reduced Cooper pair density indicating that the vortex velocity varies non-monotonically along the channels. The bright spots describe the regions where the vortices slow down or even stop momentarily, giving rise to vortex crowding (see Supplementary Movies 5 to 8). Similar features of $B_z(x, y)$ along the channels are observed in Figs. 2i-p. Such vortex "traffic jams" can be understood as follows. A vortex penetrating from the left edge slows down as it moves along the channel since the driving current $J(x, y)$ decreases across the strip. The subsequent penetrating vortices move along the same trajectory, causing jamming in the regions where vortices slow down. The resulting mutual repulsion of vortices either pushes them further along the channel, where vortices speed up due to attraction to the right edge of the strip, or causes bifurcation of the channel into branches.

The fact that there is a single vortex entry point and several exit points necessitates that the penetration frequency per stem is higher than the exit frequency per channel. Figures 5d,f show snapshots of vortex motion at different applied currents, with arrows proportional to the instantaneous velocities of vortices right after penetration of a new vortex. We find that the periodically-entering vortices take alternating routes at the bifurcation points due to interactions with other vortices, which slow down further after the bifurcation (see Supplementary Movies 5 to 8).

Figure 5g presents the calculated vortex velocity $v(x)$ along the stems in Figs. 5d and 5f juxtaposed with the experimental data. The decreasing vortex velocity follows the drop in $J(x)$ from the left edge of the constriction, $v(x) = \phi_0 J(x)/\eta$, showing no significant velocity dependence of $\eta(v)$ at these values of $v$. This allows us to extrapolate the experimental $v(x)$ to $x \simeq \xi$ at the entrance point, where the velocity is maximal, $v(\xi) \simeq J_d \Phi_0/\eta = 24$ km/s, assuming a constant $\eta(v)$.

Using the current density $J(x)$ obtained from the simulations and $v(x)$ extracted from our experimental data, we obtain the vortex viscous drag coefficient $\eta = 2.6 \times 10^{-8}$ kgm$^{-1}$s$^{-1}$. This value of $\eta$ is of the order of $\eta_0 \simeq \phi_0^2/2\pi\xi^2\rho_n = 10^{-8}$ kgm$^{-1}$s$^{-1}$ of the Bardeen-Stephen model, which indicates no excessive changes in the structure of the Abrikosov vortex core even at velocities of the order of 10 km/s. This conclusion is corroborated by our TDGL simulations in Fig. 5 which reproduce the channel bifurcations due to vortex repulsion and the variation of $B_z(x,y)$ along the channels due to disorder and interactions induced variations in vortex velocity. The totality of our SOT and TDGL results indicate that vortices maintain their integrity as stable topological defects even at the observed extreme velocities for which the magnetic field of a moving vortex does not deviate substantially from that of a stationary Abrikosov vortex. In particular, we have observed no evidence of the transition of Abrikosov vortices into Josephson-like phase slip lines[48,49] extending across the bridge.

Now we turn to the numerical study of even faster vortices, beyond our experimentally accessible range of parameters, for which a significant change in the internal vortex structure is expected. For instance, nonequilibrium effects can give rise to a velocity dependence of $\eta(v)$ and to the Larkin-Ovchinnikov (LO) instability caused by diffusion of quasiparticles from the vortex core[50]. The LO instability results in jumps in the vortex velocity above $J > J_{LO} \simeq \eta_0 v_0/2\phi_0$ for which the force balance $\phi_0 J = \eta(v)v$ at $v > v_0$ is not satisfied because $\eta(v) = \eta_0/(1 + v^2/v_0^2)$ decreases with $v$ [50]. The LO or overheating instabilities[51,52], have been observed on various superconductors with $v_0$ ranging from 1 to 10 km/s [17–19].

Our TDGL calculations at twice higher current and field as compared to those shown in Fig. 5, reveal three different types of vortices described in Fig. 6. Far from the constriction region, $J$ is lower and the moving vortices (red dot in Fig 6a) have a regular, nearly isotropic shape with no wake of reduced order parameter. Closer to the constriction, a chain of vortices (marked by the black dot in Fig. 6a) is confined in a channel of reduced order parameter. These faster-moving vortices are slipstreaming one another because their velocity $v$ exceeds $a/\tau_\Delta$, where $\tau_\Delta = \pi\hbar\sqrt{1+\gamma^2\Delta^2}/8k_B(T_c - T)$ is a recovery time of the superconducting order parameter in the wake of the moving vortex, $\gamma = 2\tau_{in}\Delta_0/\hbar$, and $\tau_{in}$ is an electron-phonon inelastic scattering time. Our TDGL simulations show that these vortices, moving in channels, have elongated cores along the direction of motion, and their drag coefficient can be approximated by the LO dependence $\eta_{LO}(v) = \eta_0/(1 + v^2/v_0^2) + \eta_i$ with $\eta_i \approx 0.25\eta_0$ and $v_0 \approx \xi/\tau_\Delta \approx 20$ km/s for our sample parameters (see Methods). These anisotropic slipstreamed vortices can undergo a kinematic transition to conventional vortices upon stem bifurcation which leads to additional vortex slowdown, as marked by the blue dot in Fig. 6a.

The most radical change in the structure of moving vortices occurs in the narrowest part of the constriction, where $J$ is maximal. Here a channel with a significant reduction of the mean superfluid density appears, in which ultrafast vortices (green dot in Fig. 6a) are moving with velocities that are 3-5 times higher than the speed of slipstreamed vortices, as shown in Fig. 6c. The ultrafast vortices in the central channel can be regarded as Josephson or mixed Abrikosov-Josephson vortices similar to vortices at grain boundaries[53,54], in high-critical-current planar junctions[55], or S/S'/S weak links[56]. The TDGL results shown in Fig. 6c suggest that Josephson-like vortices in these channels can move with velocities as high as ~100 km/s, because the viscous drag coefficient $\eta(v)$ in the channel is reduced to just a few percent of $\eta_0$ due to strongly elongated and overlapping vortex cores. Spatial modulation of the order parameter between these vortices is rather weak and effectively the channel behaves as a self-induced Josephson junction, which appears without materials weak links. Similar flux channels in thin films were previously interpreted in terms of phase slip lines[48,49]. In the case of strong suppression of the order parameter and weak repulsion of Josephson vortices which extend over lengths exceeding $\Lambda = \lambda^2/d$, the channel does not bifurcate as shown in Fig. 6a and the magnetic field along the channel is nearly uniform. This feature of Josephson-like vortices appears inconsistent with our SOT observations of vortex channels which always bifurcate and show noticeable variations of $B_z(x, y)$ along the channels. The SOT results thus indicate an essential effect of intervortex repulsions and weak suppression of the order parameter along the channel, consistent with the dynamics of Abrikosov vortices shown in Fig. 5.

Another interesting SOT observation shown in Figs. 2n, 2o and 2p is the nucleation of additional stems of vortices as current increases. This effect can be understood as follows. The first stem appears at $I = I_c$ as the local current density $J_s$ at the edge of the constriction reaches $J_d$. As $I$ increases above $I_c$, vortices start penetrating at the narrowest part of the constriction in such a way that a counterflow of circulating currents produced by a chain of vortices moving in the central channel maintains the current density $J_s(y) < J_d$ everywhere along the curved edge of the film except for the vortex entry point. This condition defines the spacing $a(I)$ between the vortices in the chain. However, above a certain current $I > I_1$, a single chain of vortices can no longer maintain $J_s(y) < J_d$ along the rest of the constriction edge, leading to nucleation of an additional stem as seen in Fig. 2n. Our calculation presented in Supplementary Note 6 and Supplementary Figs. 5 and 6 show that the second stem appears at the current $I_1 = I_c[1 + (5\sqrt{3}\xi/d)^{1/2}\lambda/R]$ which depends on the radius of curvature $R$ of the constriction. For $\xi = 46$ nm, $d = 75$ nm, $\lambda = 96$ nm and $R = 2$ μm, we obtain $I_1 = 1.11 I_c$ in good agreement with the observed $I_1 = 1.09 I_c$ at $B_a = 2.7$ mT in Fig. 3b. In addition, the edge roughness can affect the location and the dynamics of stem evolution, favoring stem nucleation at points of local edge protuberances (Supplementary Note 6). We have incorporated the actual details of the edge shape of our sample derived from the SEM image into our TDGL simulations resulting in the observed asymmetry between the vortex channels in the upper and lower parts of Figs. 6a and 6b.

As the magnetic field increases, the width of the vortex-free region near the edges and vortex velocities decrease. Figure 2 shows how dissipative vortex structures evolve from a few mesoscopic chains and branches sustaining extremely high vortex velocities at low field (Figs. 2j,n) to a multi-chain structure with much lower vortex velocities at higher fields (Figs. 2l,p). Remarkably, the vortex channeling is preserved even at high fields that would usually be associated with the conventional flux flow of an Abrikosov lattice. The dynamic structure revealed in Fig. 2p, in which vortices move in parallel channels, appears consistent with the predictions of the moving Bragg glass theory[57], thus providing microscopic evidence with a single vortex resolution for the existence of this dynamic phase.

In conclusion, this work uncovers the rich physics of ultrafast vortices in superconducting films and offers a broad outlook for further experimental and theoretical investigations. By proper sample design and improved heat removal it should be possible to reach even higher velocities for investigation of non-equilibrium instabilities[17,18,50–52,58]. Our detection of vortices moving at velocities of up to 20 km/s, significantly faster than previously reported, strengthens the recently renewed appeal of vortex-based cryogenic electronics[59]. The observed frequencies of penetration of vortices in excess of 10 GHz may be pushed to the much technologically desired THz gap in the case of dynamic Abrikosov-Josephson vortex phases. This work shows that the SOT technique can address some outstanding problems of nonequilibrium superconductivity and ultrafast vortices in type II superconductors as well as dynamics of the intermediate state in type I superconductors on the nanoscale. These issues can also be essential for further development of superconducting electronics, opening new challenges for theories and experiments in the yet unexplored range of very high electromagnetic fields and currents.

**Methods**

**Scanning SQUID-on-tip microscopy.** The SOT was fabricated using self-aligned three-step thermal deposition of Pb at cryogenic temperatures, as described previously[30,60,61]. Supplementary Figure 7 shows the measured quantum interference pattern of the SOT used for this work with an effective diameter of 228 nm, 135 µA critical current at zero field, and white flux noise down to 270 $n\Phi_0Hz^{-1/2}$ at frequencies above a few hundred Hz. The slightly asymmetric structure of the SOT gives rise to a shift of the interference pattern resulting in good sensitivity even at zero applied field with flux noise of 1.6 $\mu\Phi_0Hz^{-1/2}$. All the measurements were performed at 4.2 K in He exchange gas of ~1 mbar.

**Sample fabrication.** A 75 nm-thick Pb film was deposited by thermal evaporation onto a Si substrate cooled to liquid nitrogen temperature in order to reduce the high surface mobility of Pb atoms and limit island growth. The base pressure was $2.2\times10^{-7}$ Torr and the deposition rate was 0.6 nm/s. A protective layer of 7 nm of Ge was deposited in situ to prevent oxidation of the Pb. The sample was patterned using a standard lift-off lithographic process. The film was characterized by Scanning Electron Microscopy (SEM) and Atomic Force Microscopy (AFM) as described in Supplementary Note 2 and Supplementary Figs. 1 and 2. Further characterization of films grown under the same conditions are described in Ref. [11].

**Vortex state preparation.** To prepare the initial field-cooled vortex state a current of a few tens of mA was applied to the microbridge, heating it to above $T_c$. The current was then turned off and the film was field-cooled in the desired applied magnetic field. Supplementary Fig. 3 shows six 12×12 µm² scans of the microbridge field cooled in 0.3, 0.6, 1.5, 2.7, 5.4 and 12 mT.

**Numerical simulations.** The numerical simulations of the kinematic vortex states were performed using the generalized TDGL model for a gapped dirty superconductor[9], where the equation for the complex order parameter $\Psi(\mathbf{r},t) = \Delta(\mathbf{r},t)e^{-i\theta(\mathbf{r},t)}$,

$$\frac{u}{\sqrt{1+\gamma^2|\Psi|^2}}\left(\frac{\partial}{\partial t} + i\varphi + \frac{\gamma^2}{2}\frac{\partial|\Psi|^2}{\partial t}\right)\Psi = (\boldsymbol{\nabla} - i\mathbf{A})^2\Psi + (1-|\Psi|^2)\Psi, \qquad (1)$$

is solved self-consistently with the equation for the electrostatic potential $\varphi$:

$$\nabla^2 \varphi = \nabla(Im\{\Psi^*(\nabla - i\mathbf{A})\Psi\}). \tag{2}$$

These equations are given in dimensionless form. The distances are expressed in units of the coherence length $\xi$, the time in units of $\tau_{GL} = \pi\hbar/8k_BT_c(1 - T/T_c)u$ where $u = 5.79$ is given by the TDGL theory in the dirty limit[9]. The complex order parameter $\Psi$ is given in units of $\Delta_0 = 4k_BT_c u^{1/2}\sqrt{1 - T/T_c}/\pi$, and electrostatic potential $\varphi$ in units $\varphi_0 = \hbar/e^*\tau_{GL}$. Vector potential $\mathbf{A}$ is scaled by $A_0 = \phi_0/2\pi\xi$ and current density $j$ is given in units of $j_{GL} = \sigma_n\varphi_0/\xi$, where $\sigma_n = 1/\rho_n$ is the normal state conductivity. Parameter $\gamma = 2\tau_{in}\Delta_0/\hbar$ contains the influence of the inelastic phonon-electron scattering time $\tau_{in}$ on the dynamics of the superconducting condensate.

The simulations were implemented using a finite difference method, on a Cartesian map with a dense grid spacing of $0.1\xi$, where we reproduced the geometry of the experimental specimen based on the SEM image. Due to memory and time constraints of the self-consistent calculation, the actual size of the simulated sample was taken twice smaller than the experimental specimen, still making a formidable numerical effort on an approximately $10^3 \times 10^3$ 2D spatial mesh, where parameter $\gamma = 100$ was taken as an order of magnitude estimate for Pb (correspondingly lowering the time step in the used implicit Crank-Nicolson method). Equation (2) was solved by a spectral Fourier procedure. For such a demanding numerical task we employed the GPU-based parallel computation scheme[62]. The gauge selected for the system of Eqs. 1 and 2 is $\nabla\mathbf{A} = 0$. The boundary conditions used at the superconductor-vacuum boundary were $\mathbf{n} \cdot (\nabla - i\mathbf{A})\Psi = 0$ and $\mathbf{n} \cdot \nabla\varphi = 0$ ($\mathbf{n}$ being the unit vector perpendicular to the boundary). The current was applied through normal-metal contacts sufficiently far from the constricted area, ensuring that applied current is fully transformed into normal current there ($\mathbf{n} \cdot \sigma_n\nabla\varphi = J$).

We emphasize that the TDGL theory was used here as the best, currently available, computational tool to qualitatively address the essential physics of the interplay of inhomogeneous transport and magnetization current densities which drive the vortex matter in the bridge and produce the branching flux-flow instabilities and the striking changes in the core structure of fast moving vortices. We believe that this approach captures qualitative features of our SOT observations, although it can hardly give reliable numerical estimates of terminal vortex velocities as discussed in Supplementary Note 1. We also made more physically transparent analytic estimates of vortex confinement and buckling instabilities of vortex chains using the thermal diffusion equation and the London theory (Supplementary Notes 4 to 6).

**Data availability**

The datasets generated during and/or analysed during the current study are available from the corresponding authors on reasonable request.

**Acknowledgments**

We would like to thank M.L. Rappaport for fruitful discussions and technical support. This work was supported by the US-Israel Binational Science Foundation (BSF) grant No. 2014155 and the Israel Science Foundation grant No. 132/14. AG was also supported by the United States Department of Energy under Grant No. DE-SC0010081. M.V.M. acknowledges support from Research Foundation-Flanders (FWO). The work of Ž.L.J. and A.V.S. was partially supported by "Mandat d'Impulsion Scientifique" MIS F.4527.13 of the F.R.S.-FNRS. This work benefited from the support of COST action MP-1201.

**Author contributions**

MVM carried out the theoretical analysis and the numerical simulations. AG, YA, MVM, EZ, LE, and ŽLJ wrote the paper with contributions from all authors.

The authors declare no competing financial interests.

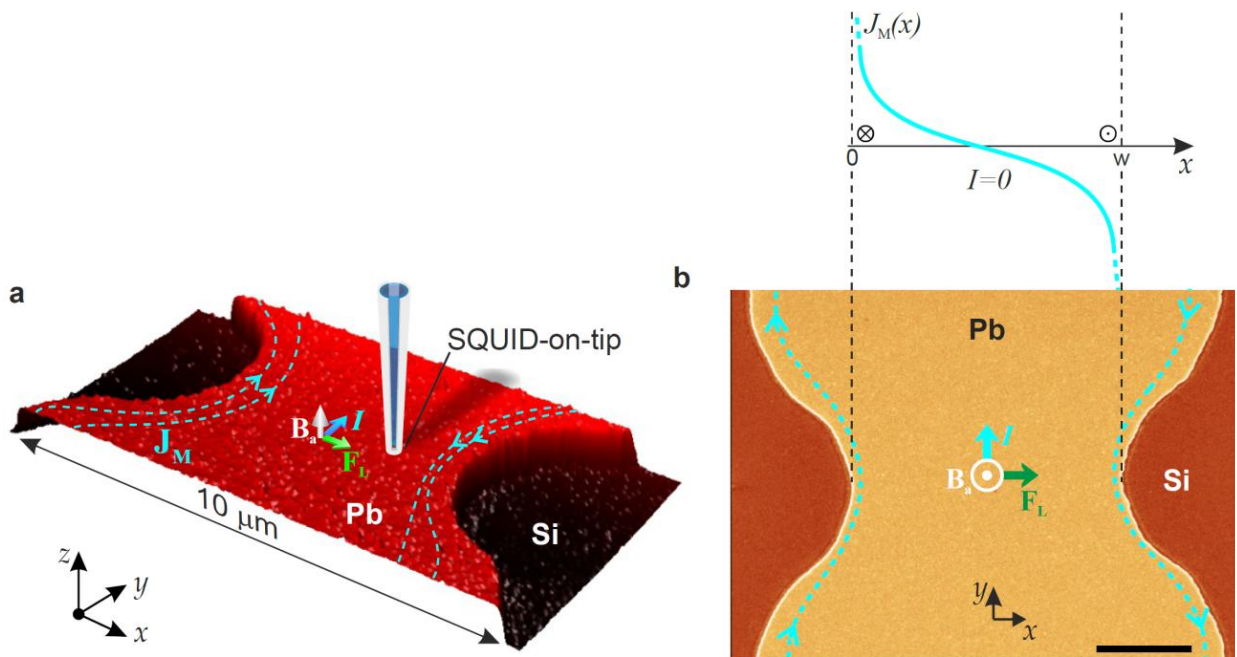

**Figure 1 | Pb thin film sample and the experimental setup.** (**a**) 3D representation of a 10×5 μm² AFM scan of the sample of 75 nm-thick Pb film patterned into a 10 μm-wide strip with a 5.7 μm wide constriction. Indicated are the directions of the applied magnetic field $B_a$, current $I$, the Lorentz force acting on vortices $F_L$, and the screening (Meissner) current density $J_M$ that is maximal along the edges. (**b**) SEM image of the same sample with corresponding distribution of the Meissner current $J_M(x)$ across the construction in absence of vortices and applied current. Scale bar is 2 μm.

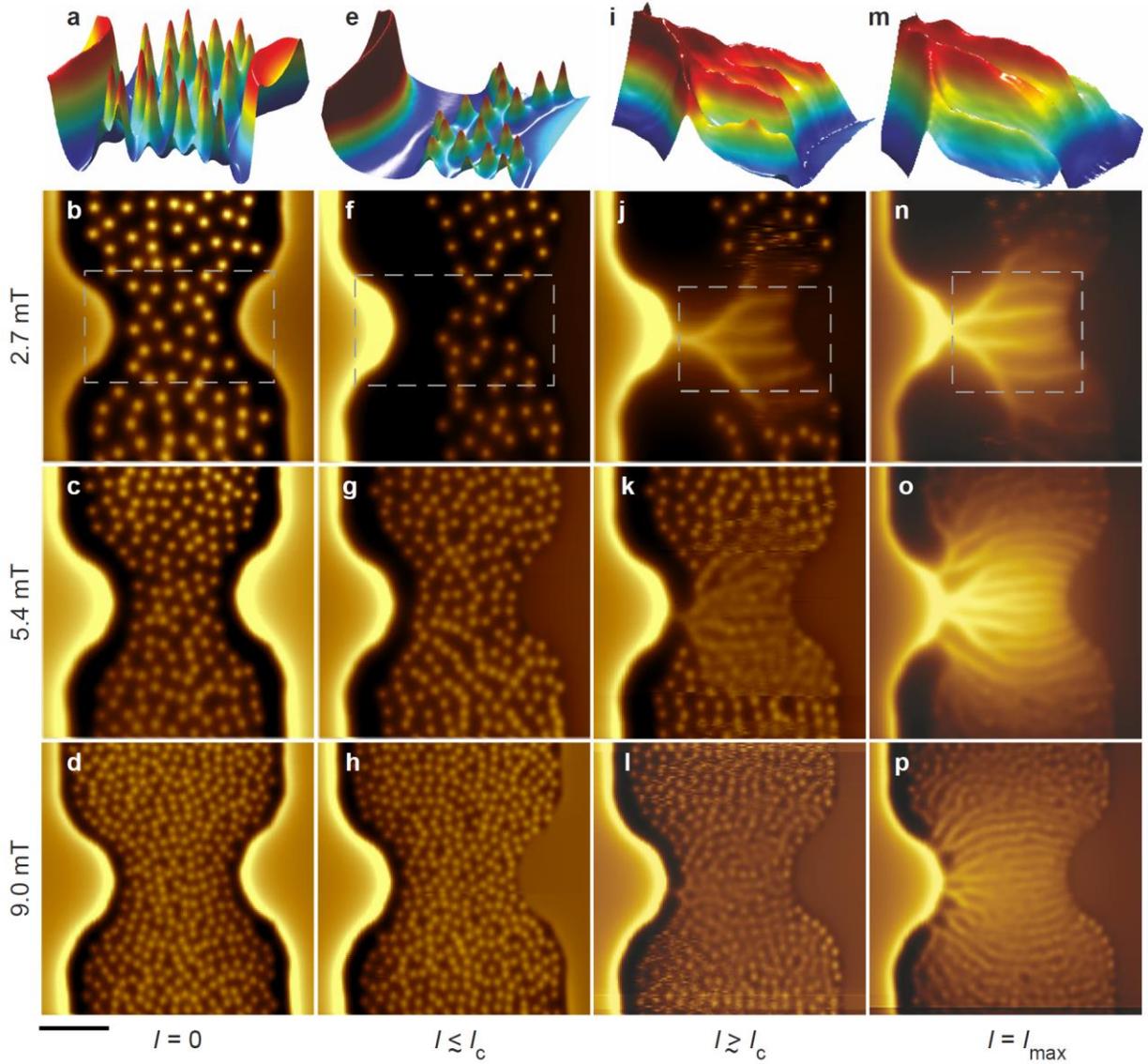

**Figure 2 | Magnetic imaging of stationary and fast moving vortices in Pb film at 4.2 K**. (a-d) $B_z(x,y)$ SQUID-on-tip images of vortex configurations at $I = 0$ for different values of applied field $B_a = 2.7$ (a,b), 5.4 (c), and 9.0 mT (d). (**e-h**) Images acquired at the verge of vortex motion at $I \lesssim I_c$, at $B_a = 2.7$ mT and $I = 16$ mA (e,f), $B_a = 5.4$ mT, $I = 12.2$ mA (g), and $B_a = 9.0$ mT, $I = 6.0$ mA (h). (**i-l**) Images of onset of vortex flow at $I \gtrsim I_c$ at $B_a = 2.7$ mT, $I = 18.9$ mA (i,j), $B_a = 5.4$ mT, $I = 12.4$ mA (k), and $B_a = 9.0$ mT, $I = 9.1$ mA (l). (**m-p**) Vortex flow patterns at the highest sustainable current with $B_a = 2.7$ mT, $I = 20.9$ mA (m,n), $B_a = 5.4$ mT, $I = 16.2$ mA (o), and $B_a = 9.0$ mT, $I = 11.8$ mA (p). The color scale represents the out-of-plane field $B_z(x,y)$ with span of 1.8 (b), 2.5 (c), 3.0 (d), 2.9 (f), 3.2 (g), 3.4 (h), 3.1 (j), 3.4 (k), 3.4 (l), 3.1 (n), 3.6 (o), and 2.8 mT (p). All 2D images are 12×12 μm², pixel size 40 nm, and acquisition time 240 s/image. The top row shows zoomed-in 3D representation of $B_z(x,y)$ in the corresponding dashed areas marked in the second row. See Supplementary Movies 1-4 for full set of images.

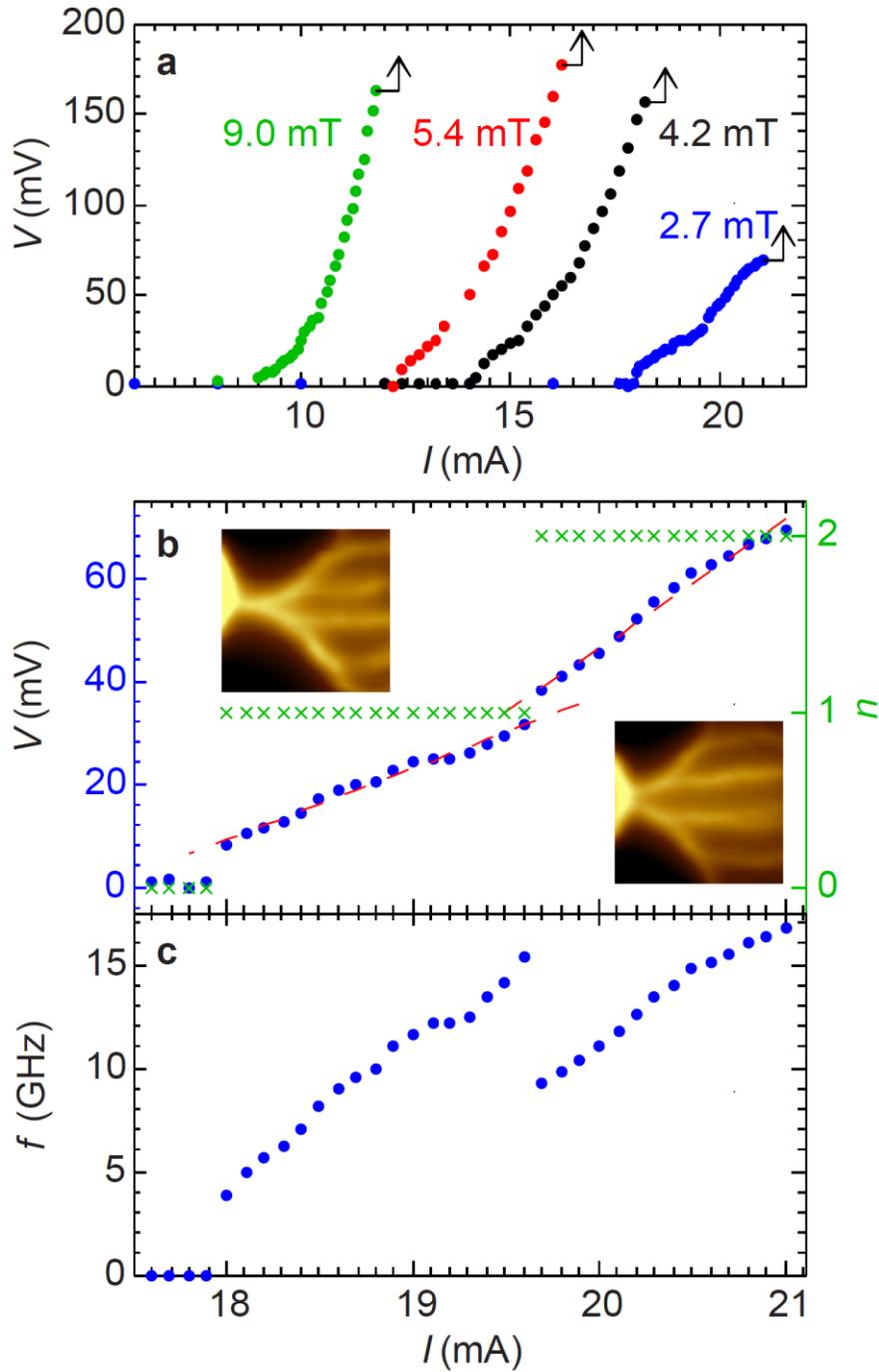

**Figure 3 | Current-voltage-frequency characterization of vortex penetration.** (**a**) Voltage $V$ across the microbridge as a function of current $I$ for various indicated fields. (**b**) Voltage across the bridge (blue) and the number of vortex stems $n$ (green) as a function of current at $B_a = 2.7$ mT. The red dashed lines are linear fits with $dV/dI = 13.9$ m$\Omega$ in the single stem and 25.1 m$\Omega$ in double stem regions. The insets show zoomed-in SOT images of single stem and double stem vortex flow. (**c**) Vortex penetration rate $f$ per stem *vs.* current $I$.

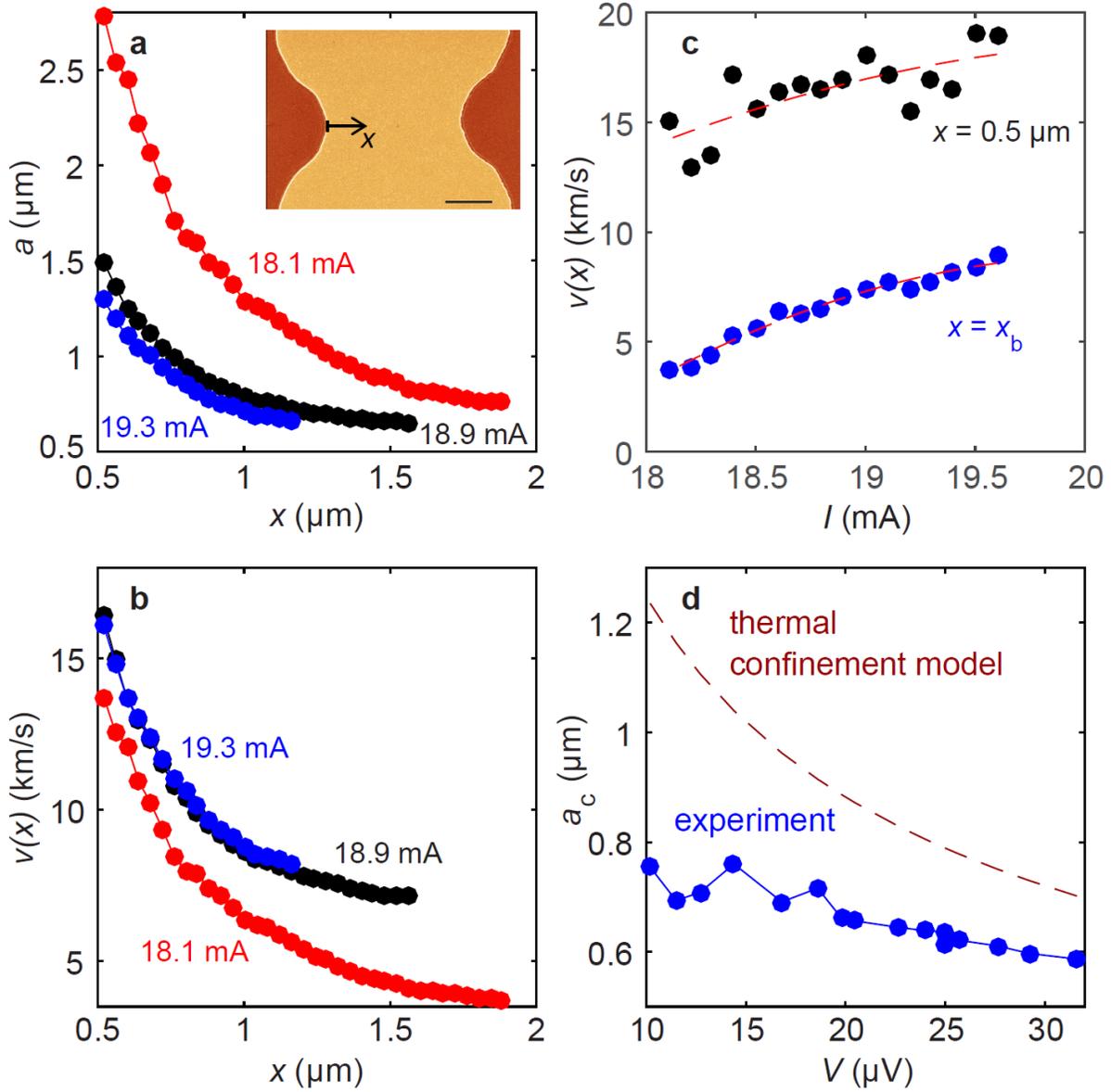

**Figure 4 | Vortex velocities and spacing along the entrance stem channel.** (a) Spacing between successive vortices $a(x)$ along the stem from $x = 0.5$ μm up to the bifurcation point $x_b$ at $B_a = 2.7$ mT at various indicated currents. Inset: SEM image of the sample with marked $x$ axis. (b) Corresponding vortex velocities $v(x)$ along the stem from $x = 0.5$ μm up to the bifurcation point $x_b$. (c) The vortex velocity *vs.* current at $x = 0.5$ μm (black) and at $x_b$ (blue). The dashed lines are guides to the eye. (d) Vortex spacing $a_c$ at the bifurcation point $x_b$ *vs.* the voltage $V$ across the microbridge (blue), compared to the theoretical estimate (dashed) based on the thermal confinement model.

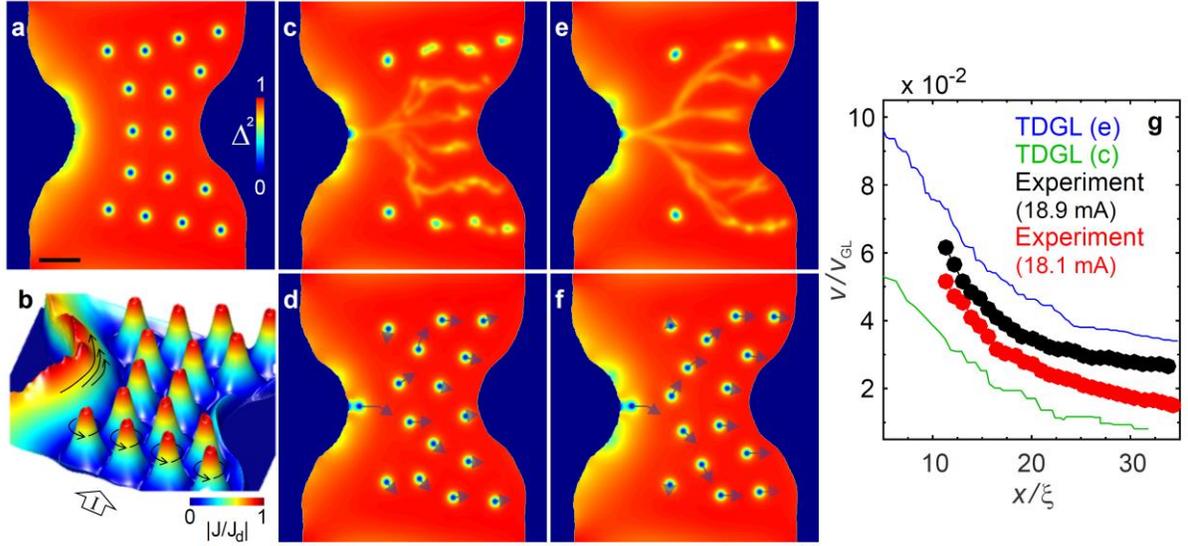

**Figure 5 | TDGL simulations of stationary and fast moving vortices at the experimentally accessible velocities.** (**a**) Calculated Cooper-pair density $\Delta^2(x,y)$ of a stationary vortex configuration at applied current density and magnetic field corresponding to the experimental conditions in Fig. 2f. (**b**) Corresponding distribution of the supercurrent density $|J(x,y)/J_d|$ in the sample showing edge currents in the constriction reaching $J_d$ at the verge of vortex penetration. The black arrows point to the local direction of the current. (**c**) Time-average of the Cooper-pair density over $5 \times 10^4 \tau_{GL}$ at $I = 1.05 I_c$, revealing branching vortex trajectories coexisting with adjacent stationary vortices. (**d**) Snapshot of moving vortices in (c) with arrows denoting the relative displacement of each vortex following an entry of a new vortex into the sample. (**e,f**) Same as (c,d) but at highest applied current before an additional stem is formed. (**g**) Experimental vortex velocity along the stem for $B_a = 2.7$ mT and indicated applied currents with the TDGL data from (d) and (f) in normalized units (scaled to $v_{GL} = \xi/\tau_{GL}$). The animation of the vortex flow dynamics corresponding to (e,f) is presented in Supplementary Videos 5 and 6. The scale bar in (a) is 20 $\xi$.

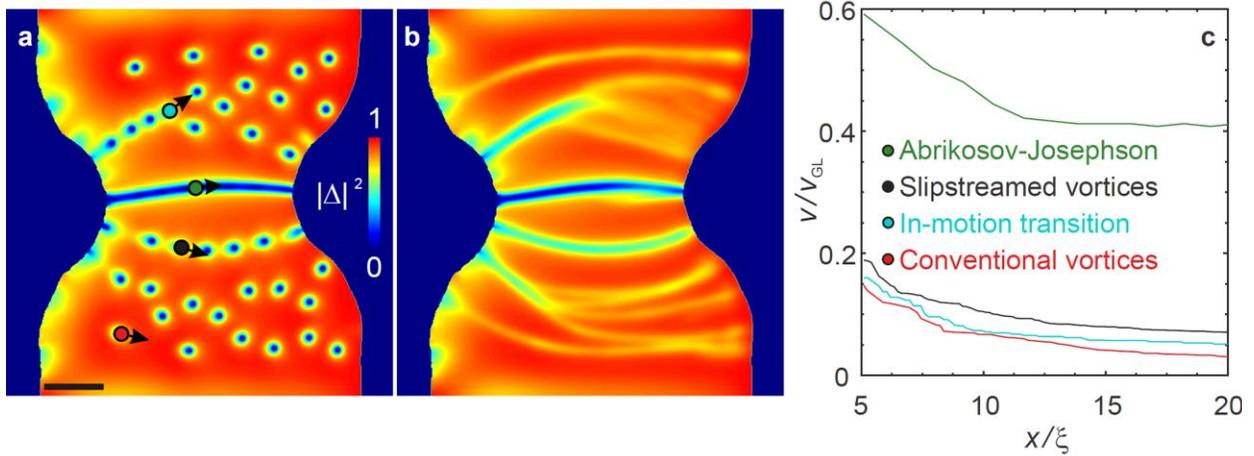

**Figure 6 | Different morphologies of ultra-fast vortices at velocities significantly higher than in our experiment.** (**a**,**b**) A snapshot (**a**) and time-averaged Cooper-pair density $\Delta^2(x,y)$ (**b**) as in Fig. 5, but for twice higher applied field and twice the current. Three vortex phase are found with distinctly different core structure, level of quasiparticle tailgating, velocities and resulting kinematic trajectories (see text and Supplementary Movies 7 and 8), namely the extremely fast Abrikosov-Josephson vortices (marked by green dot), the ultrafast slipstreamed vortices (black dot), and conventional Abrikosov moving vortices (red dot). (**c**) Spatial profiles of vortex velocities $v(x)$ (scaled to $v_{\text{GL}} = \xi/\tau_{\text{GL}}$) for the three main vortex phases, and for one detected branch of vortices going through an in-motion transition (dynamic transition from slipstreamed vortices to conventional Abrikosov vortices, identified by blue dot in (a)). The scale bar in (a) is 20 $\xi$.

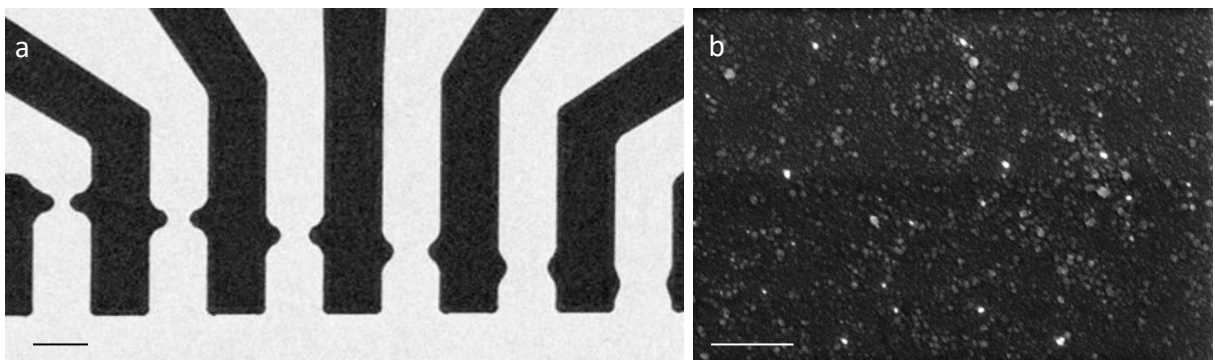

**Supplementary Figure 1 | SEM images of the sample.** (**a**) Large scale image showing the Pb film (light gray) on a Si substrate (dark). Six 10 μm wide microbridges with constrictions of different widths from 3 to 8 μm. Current can be applied independently to each of the bridges and voltage was measured at contacts outside the field of view. The scale bar is 10 μm (**b**) Higher magnification image of the surface of the Ge-capped Pb film showing grains with a typical diameter of a few tens of nm. The scale bar is 400 nm

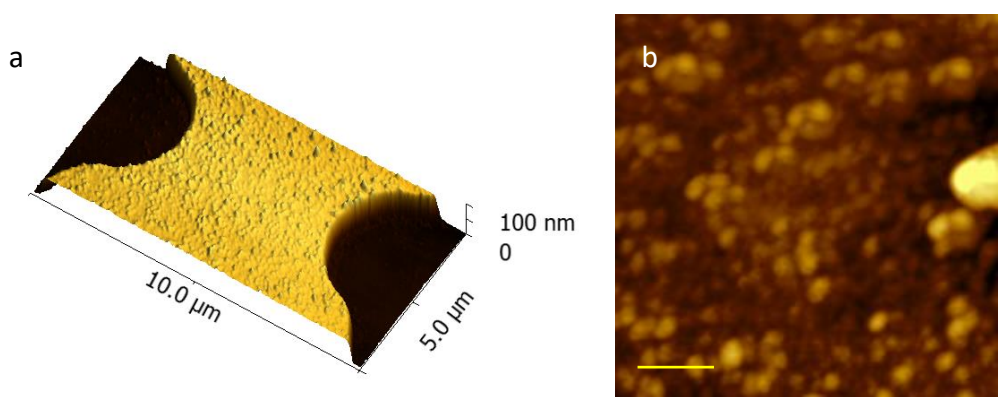

**Supplementary Figure 2 | AFM topography of the Pb sample.** (**a**) A 3D representation of the 10 μm-wide microbridge with 5.7 μm-wide constriction region. The film thickness is 82 nm; 75 nm of Pb capped by 7 nm of Ge. (**b**) Zoomed-in image of 1×1 μm$^2$ of the film surface revealing a granular structure with typical grain size of a few tens of nm. The color span is 18 nm. The scale bar is 200 nm

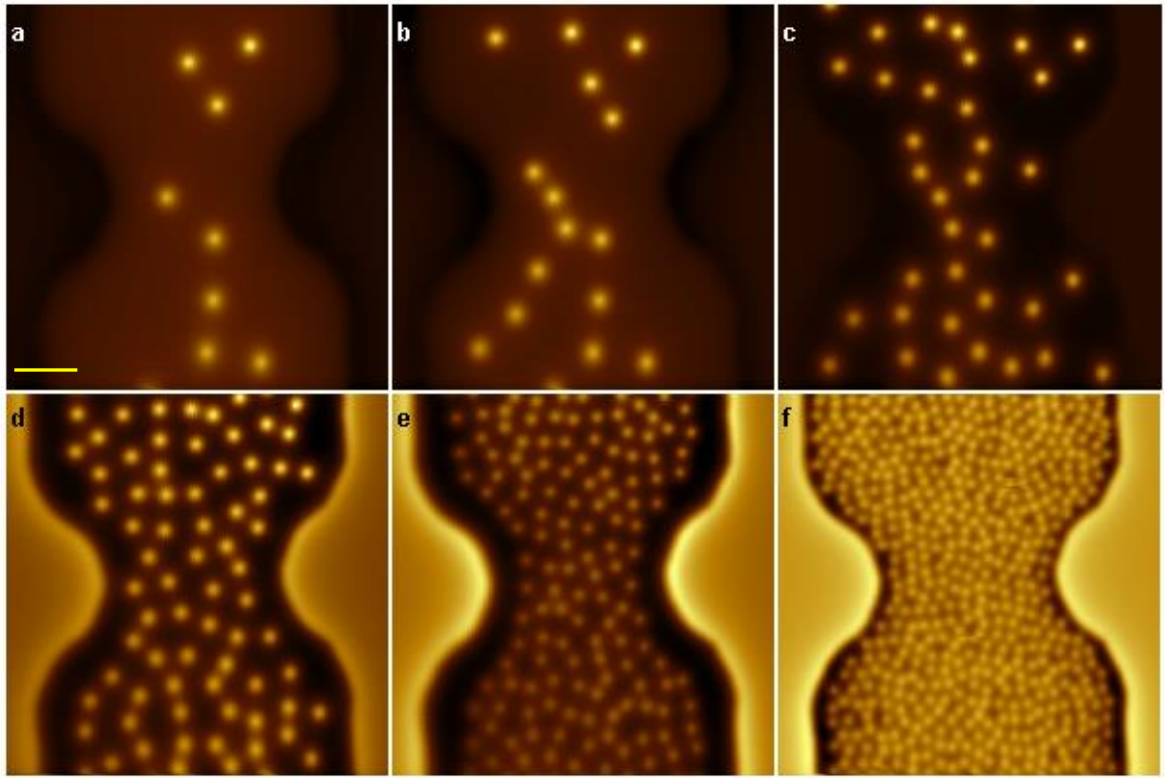

**Supplementary Figure 3 | magnetic imaging of vortex configurations after field cooling.** 12×12 μm² images showing vortices in the microbridge with the 5.7 μm-wide constriction after field cooling in fields of 0.3 (**a**), 0.6 (**b**), 1.5 (**c**), 2.7 (**d**), 5.4 (**e**), and 12 mT (**f**). The SOT scanning speed was 30 μm/s and the pixel size is 40x40 nm². The scale bar is 2 μm

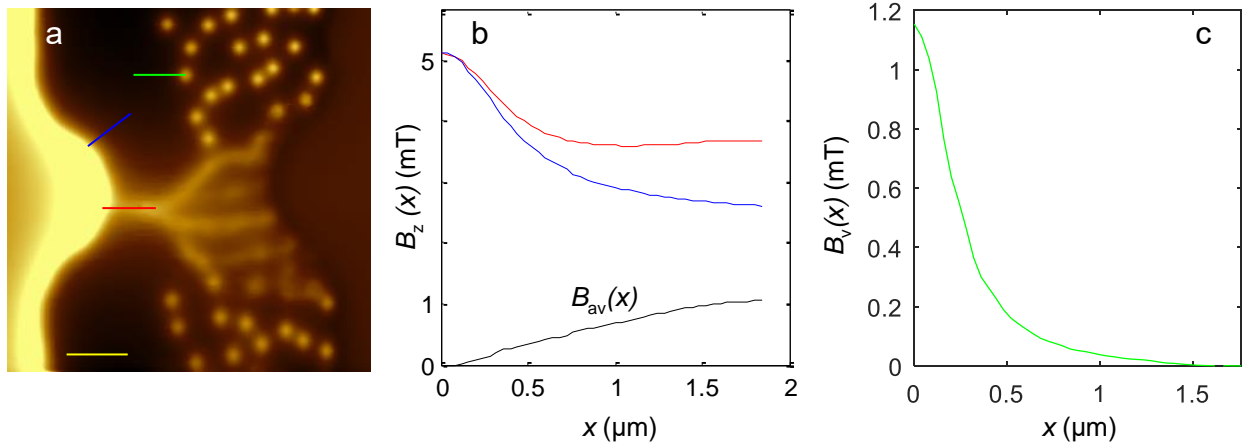

**Supplementary Figure 4 | Analysis of field profiles along an isolated vortex and vortex stem.** (**a**) $B_z(x,y)$ SOT image of the Pb microbridge at $B_a = 2.7$ mT and $I = 18.3$ mA. Scan area is 12×12 μm² and the field span is 3.1 mT. (**b**) $B_z(x)$ profiles along the lines marked in (a): Red – profile along the stem from the sample edge to the first bifurcation point $x_b$. Blue – background signal due to the Oersted field created by the edge currents. Black – $B_{av}(x)$ along the vortex stem after background subtraction. (**c**) Field profile along an isolated vortex $B_v(x)$ from the vortex center into the Meissner region (green line in (a)) after background subtraction. The scale bar is 2 μm

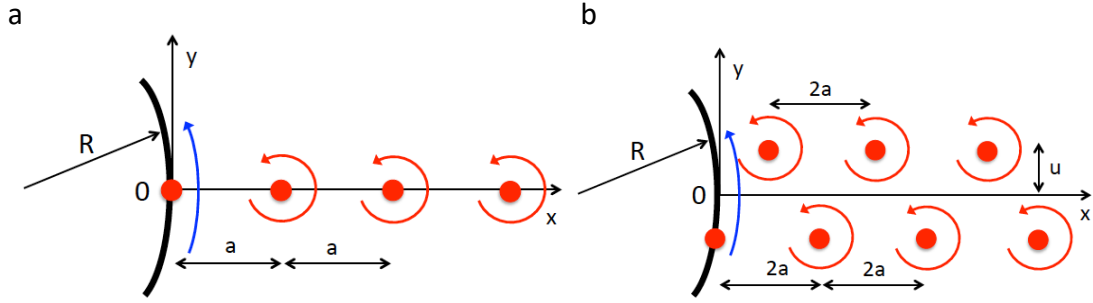

**Supplementary Figure 5 | Current flow in vortex stems.** (**a**) A single vortex chain of period $a$ at the moment of penetration of a new vortex at $(x, y) = (0,0)$. The vortex and the Meissner currents are indicated by red and blue arrows, respectively. (**b**) Two vortex stems spaced by $l = 2u$ with anti-correlated vortex arrangement at the moment of penetration of a new vortex at $y = -u$.

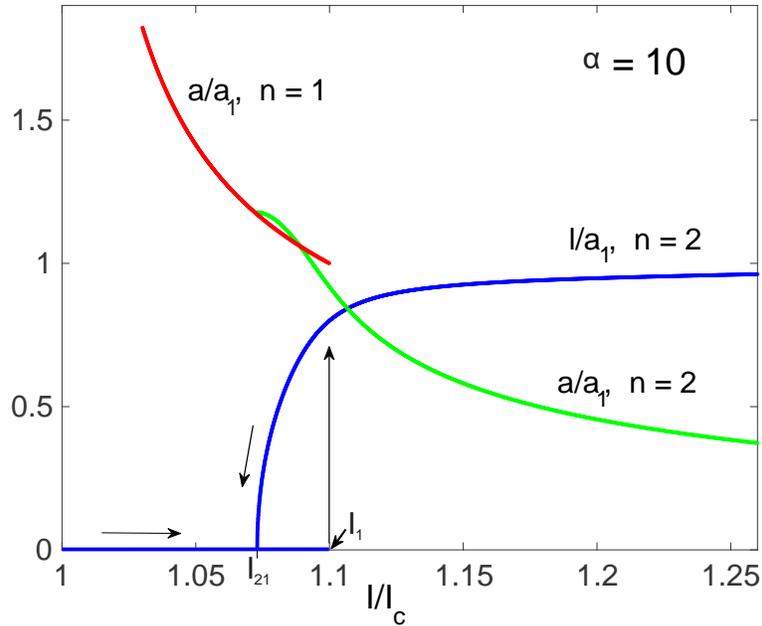

**Supplementary Figure 6 | Inter-vortex distance $a$ and inter-chain spacing $l$ for single and double chain vs. applied current $I$.** The curves $a(I)/a_1$ (red and green) and $l(I)/a_1$ (blue) were calculated from Supplementary Eqs. 34 and 35 for $\alpha = (R/\lambda)(d/5\sqrt{3}\xi)^{1/2} = 10$. The inter-vortex spacing $a(I)$ for a single chain ($n = 1$, red) decreases with $I$ up to $a = a_1$ at $I = I_1$ above which the chain splits into two chains ($n = 2$, green). For $I > I_1$ the spacing between the chains $l(I)$ has a jump followed by a continuous increase. At $I = I_1$, $a(I)$ jumps from $n = 1$ curve (red) to $n = 2$ curve (green) and then decreases continuously until the next splitting transition. This chain-splitting transition is hysteretic: as $I$ decreases, $l(I)$ decreases continuously and vanishes at $I = I_{21} < I_1$. For the presented case of $\alpha = 10$, $I_1 = 1.1 I_c$ and $I_{21} = 1.07 I_c$. For the parameters of our Pb bridge given in the text we attain $\alpha = 9$ resulting in $I_1 = 1.11 I_c$ and $I_{21} = 1.08 I_c$.

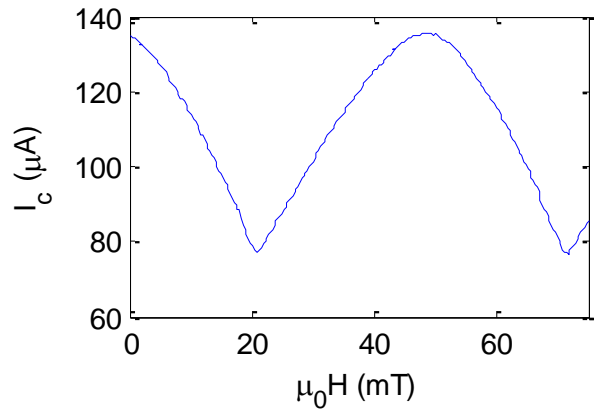

**Supplementary Figure 7 | Quantum interference pattern of the SQUID-on-tip (SOT).** The critical current $I_c$ of the SOT used in this work *vs.* the applied field. The period of 50.7 mT of the quantum interference patter corresponds to an effective diameter of 228 nm of the SOT. The asymmetric structure of the SOT resulted in a slight asymmetry and a shift in the interference pattern giving rise to finite magnetic sensitivity of the SOT down to zero applied field.

**Supplementary Note 1: TDGL description of moving vortex matter**

As described in Methods, the dynamic evolution of the complex superconducting order parameter $\Psi(\mathbf{r},t) = \Delta(\mathbf{r},t) \cdot e^{-i\theta(\mathbf{r},t)}$ at temperatures close to the superconducting critical temperature $T_c$ can be described by the generalized TDGL equation for a gapped dirty superconductor[1,2],

$$\frac{u}{\sqrt{1+\gamma^2|\Psi|^2}}\left(\frac{\partial}{\partial t} + i\varphi + \frac{\gamma^2}{2}\frac{\partial|\Psi|^2}{\partial t}\right)\Psi = (\nabla - i\mathbf{A})^2\Psi + (1-|\Psi|^2)\Psi, \quad (1)$$

coupled with the equation for electrostatic potential $\varphi$:

$$\nabla(\sigma_n \nabla \varphi) = \nabla(Im\{\Psi^*(\nabla - i\mathbf{A})\Psi\}). \quad (2)$$

These equations are written in dimensionless form as defined in the Methods. In what follows, we discuss the parameters of the Pb sample under study, and some important estimates of the observable quantities in TDGL.

Related to the observed channeling of ultrafast vortices, we first discuss the conditions for appearance of a wake of suppressed order parameter behind a rapidly moving vortex core[3,4]. Following the approach from Refs. [2,3], we exclude the phase of the order parameter $\theta(\mathbf{r},t)$ from Supplementary Eqs. (1) and (2) and obtain the following equation for the magnitude of the order parameter $\Delta(r,t)$:

$$\tau_\Delta \partial_t \Delta = \Delta - \Delta^3 + (\nabla^2 - \mathbf{Q}^2)\Delta. \quad (3)$$

where $\mathbf{Q} = \nabla\theta + \mathbf{A}$ is the gauge-invariant potential. The gap relaxation time $\tau_\Delta$ is given as

$$\tau_\Delta = \tau_{GL} u (1 + \gamma^2 \Delta^2)^{1/2}. \quad (4)$$

Here, $\gamma$ can be determined from the inelastic phonon-electron scattering time, $\tau_{in} = 2\hbar(sp_F)^2/7\pi\zeta(3)\lambda_p(k_B T)^3$ ($s$ is the sound velocity, $p_F$ the Fermi momentum, and $\lambda_p$ the dimensionless electron-phonon coupling constant), as

$$\gamma = \left(\frac{64\sqrt{u\epsilon}}{7\pi^2\zeta(3)\lambda_p}\right)\left(\frac{s}{v_F}\right)^2\left(\frac{T_c T_F^2}{T^3}\right), \quad (5)$$

where $\epsilon = 1 - T/T_c$, $T_F = p_F v_F/2k_B$ is the Fermi temperature, and $v_F$ is the Fermi velocity. Taking the available single-crystal values[5] for Pb, $T_c^{(Pb)} = 7.2$ K, $T_F^{(Pb)} = 1.1 \times 10^5$ K, $s^{(Pb)} = 2$ km/s, $v_F^{(Pb)} = 1830$ km/s, $\lambda_p^{(Pb)} = 1.55$, we obtain $\gamma(T \approx T_c) \cong 334\sqrt{\epsilon}$ and $\gamma(T = 4.2 \text{ K}) \cong 1085$. Such large values of $\gamma$ indicate that the gap relaxation time $\tau_\Delta$ is mostly determined by the electron-phonon scattering:

$$\tau_\Delta \cong \frac{\tau_{in}}{\sqrt{\epsilon u}} = \frac{8\hbar}{7\pi\zeta(3)\lambda_p k_B T \sqrt{\epsilon u}}\left(\frac{sT_F}{v_F T}\right)^2. \quad (6)$$

For the parameters listed above, Supplementary Eq. (6) yields $\tau_\Delta \approx 0.2$ ns, and $\tau_{in} \approx 0.3$ ns at 4.2 K. The wake of suppressed order parameter behind the vortex core moving with the velocity $v$ extends over the length $\Lambda_T$ which can be obtained from Supplementary Eq. (3) as:

$$4\Lambda_T = v\tau_\Delta + \sqrt{v^2 \tau_\Delta^2 + 8\xi^2}. \quad (7)$$

For $\xi = 46$ nm, $\tau_\Delta = 0.2$ ns, and $v = 5$ km/s, we obtain $\Lambda_T \cong 0.5$ μm, so this short-range mechanism may not be sufficiently effective to align vortices typically spaced by ~1-2 μm in our Pb microbridge (in relation to Fig. 4) but should become dominant at higher velocities, as is clearly observed in the dynamic vortex phases of Fig. 6.

Here we must emphasize that applicability of TDGL for gapped superconductors carries important limitations, which is why the simulations in the paper are conducted to reveal the new physics of ultrafast viscosity-dominated vortex dynamics rather than to claim that they reproduce the actual nonequilibrium properties of our specific sample. The derivation of TDGL assumes that: (1) the spatial variation of order parameter $\Psi(\mathbf{r}, t)$ at temperatures $T \to T_c$ is slow over distances $r > \xi_0$, where $\xi_0$ is the BCS coherence length, and (2) the condition of quasi-equilibrium kinetics and the neglect of higher order temporal derivatives are valid if $\Psi(\mathbf{r}, t)$ varies slowly over the inelastic phonon-electron scattering time[3], $\tau_{\text{in}}$. For a moving vortex, the characteristic variation time is the velocity divided by the core size so that TDGL is applicable if $v \ll v_c \sim \xi/\tau_{\text{in}}$. For $\xi = 46$ nm and $\tau_{in} \approx 0.3$ ns, one obtains $v_c \approx 0.2$ km/s at 4.2 K. Even though the exact parametrization of our Pb sample may differ from the one taken above, it is clear that using TDGL for calculations related to observed ultrafast-moving vortices is hardly justified, even though its results are illuminating and in line with the experimental observations. We therefore reiterate that our SOT experiments probe hitherto unexplored dynamics of very fast vortices, and at low temperatures, for which a suitable theory is yet to be developed. Such theory requires taking into account complicated nonequilibrium kinetics of superconductors coupled with strong pair-breaking effects in nonuniform distributions of superfluid density and current of moving vortex patterns.

**Supplementary Note 2: Sample characterization**

Supplementary Fig. 1a shows a SEM image of the sample. Six microbridges with constrictions of different widths (ranging from 3 to 8 µm) were patterned in the film, through which individual currents could be applied. The measurements presented here were performed on the third bridge from the right. The straight part of the bridge is 10 µm wide and the constriction is 5.7 µm wide and 5 µm long. All the transport measurements were performed in four probe configuration with voltage and current contacts outside the field of view of the SEM image. Since all the leads are significantly wider than the constriction, the current density in them is well below $J_c$ and hence no vortex flow is induced. As a result, the measured finite $V$ arises only from vortex flow in the constriction.

The zoomed-in SEM image in Supplementary Fig. 1b reveals a granular structure typical of metallic surfaces with grain size on the order of a few tens of nm. An AFM topography image of a constricted bridge is shown in 3D representation in Supplementary Fig. 2a. The overall thickness of the film was measured to be 82 nm which consists of a 75 nm-thick Pb film and a 7 nm-thick Ge capping layer. Supplementary Figure 2b is a high resolution 1×1 µm$^2$ scan of the surface showing granular structure with a typical grain size of a few tens of nm, consistent with the SEM images.

**Supplementary Note 3: Extraction of the vortex velocity along a channel**

The time-averaged field $B_{av}(x)$ along a chain of vortices separated by $a(x)$ and moving at velocity $v(x) = fa(x)$ is given by $B_{av}(x) = \int_{-\infty}^{\infty} B_v(u)du/a(x) = f\int_{-\infty}^{\infty} B_v(u)du/v(x)$ (see below), where $B_v(u)$ is the magnetic field profile of an individual vortex. By measuring $B_{av}(x)$ along the stem and $B_v(x)$ across an isolated stationary vortex, we can thus attain $a(x)$ along a single stem (Fig. 4a) and combining it with the penetration rate $f$ derived from simultaneous transport measurements (Fig. 3c), we obtain the corresponding vortex velocity $v(x)$ (Fig. 4b).

Supplementary Fig. 4a shows the measured $B_z(x, y)$ at $B_a = 2.7$ mT and $I = 18.3$ mA with a single vortex stem. The field profile along a stem is shown in Supplementary Figure 4b (red). Since the SOT image is acquired at a height of $h \sim 200$ nm above the film, it contains also the $B_z$ component of the Oersted field of the currents flowing near the edge of the strip. We subtract this contribution using the $B_z$ profile in the nearby Meissner region (blue), attaining $B_{av}(x)$, shown by the black curve in Supplementary Fig. 4b. Similarly, the field profile of a single vortex $B_v(x)$ is obtained along the green line extending from the center of the vortex into the Meissner region, as shown in Supplementary Fig. 4c. We then integrate numerically the field of a single vortex $\int B_v(u)\,du$ and repeat this procedure over several vortices in various images with observed variations of a few percent. The corresponding $a(x)$ is then derived from the ratio of $B_{av}(x)$ and $\int B_v(u)\,du$, as presented in Fig. 4 in the main text.

We note here some aspects pertaining to the validity of the above procedure.

a) The measured field profile across the stationary vortex in Supplementary Fig. 4c shows FWHM width of about 0.5 μm which is much larger than $\xi = 46$ nm and $\lambda = 96$ nm. This wide profile is a result of the magnetic size of the vortex given by the Pearl penetration depth $\Lambda = 2\lambda^2/d \cong 245$ nm further broadened by the SOT diameter of 228 nm and the SOT scanning height of about 200 nm. Under these conditions the vortices are seen by SOT as a point magnetic monopole with flux $\phi_0$. Hence the SOT images are insensitive to the elongation of the vortex core along the direction of motion, unless the core becomes longer than $\Lambda$ which is not the case in the experiment as discussed in the main text.

b) The derivation of vortex velocity does not include details of the vortex profile $B_v(x)$ itself but only the line integral of it. Since the areal integral of $B_v(x, y)$ equals $\phi_0$ regardless of the vortex structure, the line integral is significantly less sensitive to the microscopic structure of the vortex than $B_v(x)$ itself.

c) The above-described procedure is not valid very close to the edge of the constriction due to the image antivortex imposed by the boundary conditions. Consider the perpendicular component of the magnetic field $B_v(x, y, h)$ produced by the Pearl vortex spaced by $u$ from the film edge[6] at $x = 0$. To simplify the expressions we take the height above the film to be $h \gg \lambda^2/d$ resulting in

$$B_v(x, y, h) = \frac{\mu_0 \phi_0 h}{2\pi[(x-u)^2 + y^2 + h^2]^{\frac{3}{2}}} - \frac{\mu_0 \phi_0 h}{2\pi[(x+u)^2 + y^2 + h^2]^{\frac{3}{2}}}, \quad (8)$$

where the second term on the right comes from the vortex image. Let vortices appear periodically at the edge and move rapidly along $x$ so that the SOT signal $B_{av}(x, y)$ is proportional to $B_v(x, y, h)$ averaged over the instantaneous positions of vortices with the probability-density $\propto 1/v(x)$

$$B_{av}(x, y) \propto \int_0^\infty \left\{ \frac{h}{[(x-u)^2 + y^2 + h^2]^{\frac{3}{2}}} - \frac{h}{[(x+u)^2 + y^2 + h^2]^{\frac{3}{2}}} \right\} \frac{du}{v(u)}. \quad (9)$$

For a constant $v(u)$, integration of Supplementary Eq. (9) yields

$$B_{av}(x, y) \propto \frac{xh}{v(y^2 + h^2)\sqrt{x^2 + y^2 + h^2}}. \quad (10)$$

The vanishing SOT signal at the edge $x = 0$ results from the extinguishing of the vortex field by its image. Along the center of the channel ($y = 0$) the contribution of the vortex image becomes negligible for $x > h$. For this reason we ignored the experimental data at $x < 0.5$ μm.

If $v(x)$ varies slowly over the length $\sim h$, the function $v(u)$ in Supplementary Eq. (9) can be expanded around the point $x = u$, $y = 0$ where the integrand is peaked, and $v^{-1}(u) \approx v^{-1}(x) - (x - u)v'(x)/v^2(x)$, where the prime denotes the $x$-derivative. Then integration in Supplementary Eq. (9) gives the following relation:

$$B_{av}(x) \propto \frac{x}{v(x)h\sqrt{x^2+h^2}} - \frac{xhv'(x)}{v^2(x)\sqrt{x^2+h^2}(x+\sqrt{x^2+h^2})} \quad (11)$$

For $x > h$ and $v(x)$ varying slowly over the length $\sim h$, the second term on the right of Supplementary Eq. (11) is negligible and it results in the direct relation $B_{av}(x) = f\int_{-\infty}^{\infty} B_v(u)du/v(x)$ used in our data analysis where $B_v(u)$ is the field profile of an isolated vortex.

d) In addition to the derivation of the local vortex velocity presented in Fig. 4, we performed an independent evaluation of the average vortex velocity in the stem which does not rely on $B_v(x)$. It is based only on the measurement of the voltage drop on the bridge and on the average vortex distance determined by the applied field. The obtained average velocity of 15 km/s is consistent with the derived local velocities in Figs. 4b and 4c.

**Supplementary Note 4: Thermal confinement of vortex chains and buckling instability**

Long-range confinement of moving vortices can result from the Joule heating they produce. Consider the equation for the steady-state temperature $\theta(x,y) = T(x,y) - T_0$ relative to the bath temperature $T_0$ due to heating by a chain of vortices spaced by $a$ and moving with the velocity $v$:

$$\kappa\nabla^2\theta + \eta v^2 \sum_n f(x - vt - na, y) - \frac{K}{d}\theta = 0 \quad (12)$$

where $\kappa$ is the thermal conductivity, $\eta$ is the viscous drag coefficient and $K$ is the Kapitza thermal conductance between the film and the substrate[7]. The second term describes the power dissipated by a moving vortex chain and the last term describes heat transfer from the film to a substrate. Supplementary Equation (12) is written in the coordinate frame moving with vortices along the $x$-axis and describes $\theta(x,y)$ averaged over the washboard frequency $v/a$ (similar equation can be used to describe electron heating[3,8]). Here the dynamic term $Cv\partial_x\theta$ proportional to the specific heat $C$ is neglected, assuming that $v < v_T \sim \kappa/Ca$. For typical values[9] of $C = 2$ kJ/m³ K, $\kappa = 200$ W/mK and $a = 1$ μm, we have $v_T = 100$ km/s, well above the velocities measured in our SOT experiments. We also introduce the normalized weight function $f(x,y)$ which provides a smooth vortex core cutoff in the London model:

$$f(x,y) = \frac{1}{4\pi\xi^2} \exp\left(-\frac{x^2+y^2}{4\xi^2}\right) \quad (13)$$

For weak heating, $\theta(x,y) \ll T_0$, temperature dependencies of $\kappa$ and $K$ are negligible, and the solution of Supplementary Eq. (12) is given by the Fourier transform

$$\theta(x,y) = \frac{\eta v^2}{\kappa a} \sum_G \int_{-\infty}^{\infty} \frac{\exp[-(q^2+G^2)\xi^2 + iqy + iGx]dq}{2\pi(q^2+G^2+L_T^{-2})}, \quad (14)$$

where $G = 2\pi n/a$, $n = 0, \pm 1, \pm 2, \ldots$, and the thermal length $L_T$ defines a spatial scale of temperature variation along the film

$$L_T = (d\kappa/K)^{1/2}. \quad (15)$$

For typical values[9] of $\kappa = (200\text{-}300)$ W/mK, $K = 10\text{-}50$ kW/m²K at 4.2 K and $d = 50$ nm, Supplementary Eq. (15) yields $L_T = 10\text{-}30$ μm, much greater than the vortex spacing, so that variation of $\theta(x,y)$ along the vortex chain is negligible and the main contribution in Supplementary Eq. (14) comes from the term with $G = 0$ in the sum. In this case, calculation of the integral in Supplementary Eq. (14) and $|y| \gg \xi$ gives the temperature distribution across the vortex chain:

$$\theta(y) = \frac{\eta L_T v^2}{2\kappa a} exp\left(-\frac{|y|}{L_T}\right). \quad (16)$$

At $|y| \sim \xi$, the finite size of the vortex core in Supplementary Eq. (14) becomes important as it rounds the cusp in Supplementary Eq. (16) at $y = 0$, resulting in a finite curvature $\theta''(y)$ inversely proportional to $\xi$:

$$\theta''(0) = -\frac{\eta v^2}{2\sqrt{\pi}\xi a\kappa}. \quad (17)$$

Unlike $\theta(y)$ at $|y| \gg \xi$ in Supplementary Eq. (16), $\theta''(0)$ in Eq. (4.6) is independent of the Kapitza conductance $K$ and thus of details of heat transfer from the film to the substrate.

To estimate the maximum heating $\theta_m = \theta(0)$ in the moving vortex chain, we use Supplementary Eqs. (15) and (16) along with the power balance, $\eta v^2/a = IV/dw$ where $V$ is the constant voltage on the chain, $I$ is the total current and $w$ is the film width. Hence,

$$\theta_m = \frac{IV}{2w\sqrt{dK\kappa}}. \quad (18)$$

Taking here typical values of $I = 20$ mA, $V = 30$ μV, $d = 50$ nm, $w = 5$ μm, $K = 10$ kW/m²K and $\kappa = 200$ W/mK for our Pb microbridge, yields rather weak heating $\theta_m = 0.2$ K, which can hardly cause thermomagnetic branching instabilities[10,11].

However, even such weak heating can align vortices in a chain because the long-range temperature field extends over distances $L_T$ much greater than the inter-vortex spacing. As a result, a moving vortex chain produces a self-consistent "temperature well" which prevents buckling instability of repulsive vortices. Indeed, if a single vortex is shifted across the chain by $u$, it experiences the restoring thermal forces $f_T = -s^*\nabla\theta$, where $s^*(T)$ is the transport entropy per unit vortex length[12]. Using $\theta(y)$, calculated above, we obtain $f_T = -s^*\theta''(0)u$, where $\theta''(0)$ is given by Supplementary Eq. (17). Hence,

$$f_T = \frac{\eta v^2 s^* u}{2\sqrt{\pi}\xi a\kappa} \quad (19)$$

The thermally-induced confinement force $f_T$ increases as vortices move faster and get closer, so that the depth of the thermal well increases with the dissipated power $IV$. Yet $f_T$ in Supplementary Eq. (19) is independent of the thermal Kapitza conductance between the film and the substrate because $L_T \gg \xi$.

Shifting a single vortex by $u$ across the chain also causes a Lorentz force $f_m(u)$ from other vortices that pushes the vortex further away from the chain. Using the repulsive force $F_m = (\phi_0/2\pi r)^2$ between two vortices separated by $r > \lambda^2/d$ in a thin film[13], we calculate the sum of the $y$-components of the inter-vortex interaction forces per unit length $f_m = F_m/d$ acting on a vortex shifted by $u < a$:

$$f_m = \frac{\phi_0^2}{2\pi^2 d}\sum_{n=1}^{\infty}\frac{u}{(a^2 n^2 + u^2)^{3/2}} \cong \frac{\phi_0^2 \zeta(3) u}{2\pi^2 d a^3}. \quad (20)$$

The vortex chain is stable with respect to buckling distortions if $f_T > f_m$. It is convenient to express the vortex velocity in terms of the voltage $V = v\phi_0/ca$ and to write the stability condition $f_T > f_m$ in the form

$$a > a_c = \left[\frac{\zeta(3)\kappa\phi_0^4\xi}{\pi^{3/2}\eta V^2 c^2 ds^*}\right]^{1/4}. \quad (21)$$

This criterion shows that the branching instability of a vortex chain occurs in the region of the film where the current density decreases so that vortices slow down and the spacing $a(x)$ decreases below $a_c$. For a rough estimate of $a_c$, we use[12] $\eta = \phi_0^2/2\pi\xi^2\rho_n c^2$ and $s^* \sim (\phi_0/4\pi\lambda)^2 T/T_c^2$ where $\rho_n$ is the normal state resistivity:

$$a_c \sim 2\sqrt{\lambda\xi}\left[\frac{\kappa\rho_n T_c^2 \xi}{V^2 T d}\right]^{1/4} \sim 2\sqrt{\frac{\lambda\xi T_c}{V}}\left(\frac{L_N \xi}{d}\right)^{1/4}. \qquad (22)$$

In the last expression, $a_c$ was further simplified by using the Wiedemann-Frantz law $\kappa\rho_n = TL_N$ with the Lorentz number $L_N = (\pi k_B/e)^2/3 = 2.44 \cdot 10^{-8}$ W·Ω/K². In this case, $a_c$ becomes independent of the thermal materials parameters. For $V = 25$ μV, Supplementary Eq. (22) gives $a_c \sim 0.8$ μm, consistent with typical inter-vortex spacing in the branching vortex chains revealed by SOT. Supplementary Equation (22) also predicts that $a_c \propto V^{-1/2}$. $a_c$ increases as vortices slow down and become closer to each other, thus reducing the depth of the thermal well and increasing the buckling effect of inter-vortex repulsion. The dashed line in Fig. 4d is a plot of $a_c(V)$ given by Supplementary Eq. (22) using our Pb film parameters with no additional fitting parameters, showing good qualitative agreement.

**Supplementary Note 5: Effects of disorder on premature branching**

The above consideration addressed the buckling instability of a uniformly moving vortex chain with respect to infinitesimal bending distortions. In our experimental situation of $J \gg J_c$ the effect of pinning on the viscosity-dominated vortex dynamics is weak and the channel bifurcation occurs primarily as the separation between the slowing down vortices drops below $a_c$ as described above. For the sake of completeness, we evaluate here the contribution that disorder may have on distorting the thermal alignment of vortices in a chain, possibly causing premature branching instabilities even if the stability condition $a > a_c$ for a fully aligned channel is satisfied.

The correlation function of pinning-induced transverse vortex displacements $\langle u_y(x)u_y(0)\rangle \propto x$ increases with the distance along the chain, which indicates buckling instability for a vortex chain longer than the critical length $x_c$. The increase of $\langle u_y(x)u_y(0)\rangle$ with $x$, similar to the well-known result of the collective pinning theory[13], can be obtained from the dynamic equation for $u_y(x,y,t)$:

$$\eta \partial_t u_y = -\partial_y U(x - u_x, y - u_y) \qquad (23)$$

where $U(x,y)$ is a random pinning potential with zero mean $\langle U(x,y)\rangle = 0$. For rapidly moving vortices at $J \gg J_c$ where pinning induced potential is strongly suppressed, disorder can be treated perturbatively, replacing in the first approximation $U(x - u_x, y - u_y) = U(x - vt, y)$, where $v$ is the free flux flow velocity[13]. Then the correlation function becomes

$$\eta^2 \langle u_y(t)u_y(t')\rangle = \lim_{\substack{y \to 0, \\ y' \to 0}} \frac{\partial^2}{\partial y \partial y'} \int_0^t dt_1 \int_0^{t'} dt_2 \langle U(vt_1, y)U(vt_2, y')\rangle, \qquad (24)$$

where $x = vt$. We illustrate the qualitative features of $\langle u_y(x)u_y(0)\rangle$ for the Gaussian pinning correlation function:

$$\langle U(x,y)U(x',y')\rangle = U_0^2 e^{-R^2/2l^2}, \qquad (25)$$

where $R = |\mathbf{r} - \mathbf{r}'|$ and $U_0$ quantify the strength of pinning, while the correlation length $l$ depends on the interaction radius of pinning centers and their spatial correlation. Substituting Supplementary Eq. (25) and integrating yields

$$\langle u_y(t)u_y(t')\rangle = \frac{\sqrt{2\pi}U_0^2}{\eta^2 v^2 l}x. \qquad (26)$$

Supplementary Equation (26) shows that, as vortices move along the chain, their pinning-induced transverse displacements $\langle u_y^2\rangle^{1/2} \propto x^{1/2}$ increase with $x$, particularly as vortices slow down. Pinning thus provides a destabilizing mechanism which may cause premature bifurcation of vortex chains as they propagate into the film. A similar effect was observed in numerical simulations of thermomagnetic flux avalanches in thin films where random inhomogeneities of pinning can greatly increase dendritic branching of propagating flux filaments[10].

**Supplementary Note 6: Dynamics of stem nucleation in London model**

We address the mechanisms that lead to nucleation of additional vortex stems using the London theory. Suppose that thermal or other mechanisms of confinement provide effective alignment of vortices in a stem formed at the edge in the narrowest part of the bridge (see Fig. 2g). Let us now evaluate the spacing $a(I)$ between vortices in the stem, and the conditions under which such a stem becomes unstable with respect to nucleation of an additional stem. In the Meissner state of a wide thin-film bridge carrying transport current $I$ in a perpendicular magnetic field $H$, the maximum current density $J_m(y)$ flows at the edge at which the magnetization currents are parallel to transport current[14]:

$$J_m(y) \cong \frac{4I + cHw(y)}{4\pi d^{3/2}w^{1/2}(y)} \qquad (27)$$

Here we assume that the width of the microbridge $w(y)$ varies slowly over the Pearl length $\Lambda = 2\lambda^2/d$, and the width of the narrowest part of the bridge can be approximated by

$$w(y) = \left(1 + \frac{y^2}{R^2}\right)w_0, \qquad (28)$$

where $R \gg \Lambda$ is a characteristic curvature radius of the constriction (about 2 μm for our Pb microbridge). The non-dissipative Meissner state persists as long as $I < I_c$, where the critical current $I_c$ is defined by the condition that $J_m(0)$ at the edge of the narrowest part of the bridge reaches the depairing current density $J_d$ at which the barrier for penetration of vortices vanishes:

$$I_c = \left(1 - \frac{H}{H_0}\right)I_0 \qquad (29)$$

$$I_0 \sim \pi d^{3/2}w_0^{1/2}J_d, \qquad H_0 \sim 4I_0/cw_0.$$

At $I > I_c$ and $H = 0$, the Meissner current density exceeds $J_d$ in a segment $-L_m < y < L_m$ of length $2L_m = 2R(2(I/I_0 - 1))^{1/2}$ along the rim. As a result, a chain of vortices penetrates at $y = 0$ to produce current counterflow, which reduces the edge value of $J(y)$ below $J_d$. Consider for simplicity a periodic chain with a constant intervortex spacing $a(I)$ determined by the condition that $J_m(0)$ reaches $J_d$ as the last vortex in the chain closest to the edge is at the distance $a(I)$ from the edge, so a new vortex penetrates (Supplementary Figure 5).

A chain of Pearl vortices produces the following current density along the rim:

$$J_v(y) = \frac{c\phi_0}{2\pi^2}\sum_{n=1}^{\infty}\frac{na}{(a^2n^2 + y^2)^{3/2}}. \qquad (30)$$

The intervortex distance $a(I, H)$ is defined by the condition $J_m(0) - J_v(0) = J_d$ which yields:

$$a(I, H) = \pi\lambda \left(\frac{\xi\sqrt{3}}{d}\right)^{1/2} \left[\frac{I_0}{I - I_c(H)}\right]^{1/2}, \qquad (31)$$

where we used the GL expression, $J_d = c\phi_0/12\sqrt{3}\pi^2\lambda^2\xi$, assuming no defects at the edge.

A single vortex chain exists only in a certain range of currents $I_c < I < I_1$ where the net edge current density $J_s(y) = J_m(y) - J_v(y)$ is below $J_d$ everywhere along the rim except the entry point $y = 0$ where $J_s(y)$ is maximum and equals $J_d$. As $I$ increases, the region $-L_m(I) < y < L_m(I)$ expands, so the counterflow of a single vortex chain can no longer sustain $J_s(y)$ below $J_d$ everywhere along the rim. At $I = I_1$, the second derivative $\partial_{yy}J_s(y)$ at $y = 0$ changes sign and the function $J_s(y)$ has two maxima at two symmetric points $y = \pm u$ and $y = 0$ becomes a local minimum of $J_s(y)$. From the condition, $\partial_{yy}J_v(y) = \partial_{yy}J_m(y)$ at $y = 0$, we calculate $I_1$ and the spacing $a_1 = a(I_1)$ at $H = 0$:

$$I_1 = \left[1 + \frac{\lambda}{R}\left(\frac{5\sqrt{3}\xi}{d}\right)^{1/2}\right] I_0, \qquad (32)$$

$$a_1 = \pi \left(\frac{\sqrt{3}\xi}{5d}\right)^{1/4} \sqrt{\lambda R}. \qquad (33)$$

For $R = 1.5$ μm and $\lambda = 96$ nm, Supplementary Eqs. (32) and (33) yield $a_1 \approx 0.9$ μm and $I_1$ about 15% higher than $I_c$, qualitatively consistent with the experimental data shown in Fig. 3b.

At $I > I_1$, instead of a single stem, two stems are formed at $y = u$ and $y = -u$ with anti-phase vortex arrangement as shown in Supplementary Figure 5. Such a structure gives $J_s(y) < J_d$ everywhere along the rim except at the points $y = \pm u$ where new vortices enter. The next vortex enters at $y = -u$, so the condition $J_s(-u) = J_d$ and $\partial_y J_s(-u) = 0$ give two coupled equations for the period $a(I)$ within each stem and the inter-stem spacing $2u(I)$ derived using Supplementary Eq. (30):

$$J_d = \frac{I}{\pi d^{3/2} w_0^{1/2}}\left(1 - \frac{u^2}{2R^2}\right) - \frac{c\phi_0}{48da^2} - \frac{c\phi_0}{2\pi^2 d}\sum_{n=1}^{\infty}\frac{(2n-1)a}{[a^2(2n-1)^2 + 4u^2]^{3/2}} \qquad (34)$$

$$\frac{I}{\pi d^{3/2} w_0^{1/2} R^2} = \frac{3c\phi_0}{\pi^2 d}\sum_{n=1}^{\infty}\frac{(2n-1)}{[a^2(2n-1)^2 + 4u^2]^{5/2}} \qquad (35)$$

Shown in Supplementary Figure 6 are the results of numerical simulations of Eqs. (34, 35) for $\alpha = (R/\lambda)(d/5\sqrt{3}\xi)^{1/2} = 10$. The red line shows the vortex spacing $a(I)$ for a single stem, the green line shows the intra-stem vortex spacing $a(I)$ when two stems are present, and the blue line shows the inter-stem spacing $l(I)$. The essential feature of these results is a hysteretic single-to-double stem transition at $I = I_1$, when $l$ jumps from 0 to $0.7a_1$, while $a(I)$ drops from $a_1$ to $0.88a_1$. The hysteresis occurs in the range of currents $I_{21} < I < I_1$ where $I_{21} = [1 + 4(\xi/2\sqrt{3}d)^{1/2}\lambda/R]I_c$. As $I$ increases further, a cascade of splitting transitions to four and more stems occurs.

Thus, we have derived above the criterion of the absolute instability of the vortex chain located at the point of a minimum of the width $w(y)$ (at $y = 0$). According to this criterion, at the current $I_1$, the maximum in the edge current density $J_s(y) = J_m(y) - J_v(y)$ at $y = 0$ evolves into a minimum of this function, and the vortex stem splits into the two equivalent stems shifted from $y = 0$. This splitting gives rise to a two-fold decrease in the vortex penetration rate per stem, as observed in Fig. 3c.

In the above scenario, the function $J_s(y) = J_m(y) - J_v(y)$ has a global maximum at $y = 0$ if $I < I_1$. However, due to irregularities of the sample edges, $J_s(y)$ can have an additional local maximum at some point $y_0 \neq 0$ because $J_m(y)$ and $J_v(y)$ have different dependences on $y$. Consider the following simple example: an infinite strip with a small constriction in the form of two semicircles of radius $r \ll w$ centered at $y = 0$ and $x = 0, w$. If the critical current without the constriction is $I_c$, the critical current of the strip with the constriction is reduced to $\tilde{I}_c = I_c(1 - 2r/w)^{1/2} \cong (1 - r/w)I_c$. According to Supplementary Eq. (32), a second stem at the constriction appears at the current $I_1$, which can be larger than $I_c$ if $r^2 < w\lambda(5\sqrt{3}\xi/d)^{1/2}$. But at large $y$, the current density $J_v(y) \to 0$, and the edge current density $J_s(y) \approx J_m(y)$, which reaches the depairing current density $J_d$ at the current $I = I_c < I_1$. In this case $y_0 \to \infty$, and the second stem will appear $I = I_c < I_1$ far away from the constriction. If there is another small constriction or edge irregularity along the strip the second stem will be nucleated there rather than at infinity, so that the initial stem does not change its position upon appearance of the second stem.

The intervortex distances in the stems are now defined by the equations:

$$J_m(0) - J_v(0) - J_{v1}(0) = J_d \quad (36)$$

$$J_m(y_0) - J_v(y_0) - J_{v1}(y_0) = J_d, \quad (37)$$

where $J_{v1}(y)$ is the edge current density generated by the vortices in the second stem. If $y_0$ is sufficiently large, we can neglect $J_{v1}(0)$ in Supplementary Eq. (36) and $J_v(y_0)$ in Supplementary Eq. (17) so that the vortex penetration process in the two stems is decoupled, each having its own penetration rate and independent $V - I$ characteristic, and the overall characteristic of the strip will be given by the sum of the two. Since nucleation of an independent stem is associated with a jump in $V$ at $I_c$, the appearance of the second stem at $I_1$ will be accompanied by a jump in the total voltage of the strip comparable to the jump at $I_c$. This consideration shows that, if the edge of the bridge constriction has small protuberances of radius r << R, the first stem does not split into two stems which are symmetric with respect to y = 0. Instead, the first stem may remain near the narrowest point of the bridge at y = 0, whereas the second stem appears at a point where the nearest edge protuberance is.

A closer inspection of the experimental data in Figs. 3b and 3c indicates that both mechanisms described above are relevant. The two stems appear to be strongly coupled resulting in a small step upon splitting of the first stem into two stems at $I_1$ accompanied by a drop to about half in the penetration rate into each stem. On the other hand, instead of a symmetric splitting of the original stem at $y = 0$ into two stems at $y = \pm y_0$, the position of the first stem remains almost fixed and the second stem appears closely above it. This implies that even though the two stems are strongly coupled, their precise location is determined by the small local irregularities in the edge "coastline". Indeed, our lithographic process resulted in some edges roughness and in addition, the constriction edges have a shape of a rounded polygon rather than a semicircle (see SEM image in Fig. 1b). As a result, the third and fourth stems in Fig. 2o are seen to be formed at the rounded corners of the polygon. The fine details of the edge shape were incorporated in our TDGL simulations. Figures 5d and 5f clearly show enhanced suppression of the order parameter at these corners that become the nucleation points of new stems at higher currents as shown in Figs. 6a and 6b. The incorporated fine irregularities of the edges also result in the pronounced roughness of the order parameter at the left edge of the constriction where the current approaches $J_d$ as seen in Fig. 5b and Movies 5 to 8, and lead to the differences in the vortex flow patterns between the upper and lower stems in Figs. 6a and 6b.